\documentclass[12pt,a4paper,dvips]{article}
\usepackage{a4p}
\usepackage{cite,mcite}
\usepackage{graphicx,rotating}
\usepackage{physics }
\usepackage{epsfig }
\usepackage{l3_titlePH,ifthen}
\usepackage{Lep}

\journalname{Phys. Lett. B}
\date{September 3, 2004}
      
\preprint{2004-049}

\usepackage{Lep}
\Lep{2}

\def\Journal#1#2#3#4{{#1} {\bf #2} (#4) #3}

\def\NIMA{{Nucl. Instr. Meth.} A}
\def\NPB{{Nucl. Phys.} B}
\def\PLB{{Phys. Lett.}  B}
\def\PRL{Phys. Rev. Lett.}
\def\PRD{{Phys. Rev.} D}
\def\ZPC{{Z. Phys.} C}
\def\PRP{{Phys. Rep.} C}
\def\CPC{Comput. Phys. Commun.}

\newlength{\capwidth}
\def\pb{\mbox{pb$^{-1}$}}
\newcommand{\EE}{\mathrm{e}^+\mathrm{e}^-}

\newcommand{\gamgam}{\gamma \gamma^*}

\newcommand{\rocroc}{\mathrm{\rho^+\rho^-}}
\newcommand{\ro}{\mathrm{\rho^0}}
\newcommand{\roro}{\mathrm{\rho^0\rho^0}}
\newcommand{\pipi}{\mathrm{\pi^+\pi^-}}
\newcommand{\piz}{\mathrm{\pi^0}}

\newcommand{\q}{ Q^2 }
\newcommand{\mgg}{ W_{\gamma \gamma }}
\newcommand{\ptt}{ p_t^2 }

\begin{document}

\begin{titlepage}
\title{Measurement of Exclusive  {\boldmath$\rho^0\rho^0$} Production \\
in Mid-Virtuality Two-Photon Interactions at LEP}

\author{The L3 Collaboration}

\begin{abstract}
Exclusive $\roro$ production in two-photon collisions between a
quasi-real and a mid-virtuality photon is studied with data collected
at LEP at centre-of-mass energies \mbox{$183 \GeV < \sqrt{s} < 209
\GeV{}$} with a total integrated luminosity of $684.8~\pb$.  The cross
section of the process $\, \gamma \gamma^* \rightarrow \roro \,$ is
determined as a function of the photon virtuality, $\q$, and the
two-photon centre-of-mass energy, $\mgg$, in the kinematic region:
$0.2 \GeV^2 < \q < 0.85 \GeV^2$ and $1.1 \GeV < W_{\gamma\gamma} < 3
\GeV$.
\end{abstract}

\submitted

\end{titlepage}

\section {Introduction}

Recently, the L3 Collaboration measured the processes $\gamgam \to
\roro$ and $\gamgam \to \rocroc$, where one of the interacting
photons, $\gamma$, is quasi-real and the other, $\gamma ^*$, is
off-mass-shell and has a virtuality in the range $1.2 \GeV^2 < \q < 30
\GeV^2$ \cite{L3paper269, L3paper287}.  The cross sections of these
isospin-related reactions have a similar dependence on the two-photon
centre-of-mass energy, $\mgg$, and are of similar magnitude, though
the $\rocroc$ cross section is systematically higher than the $\ro\ro$
one.  These features of $\rho$ pair-production at high $\q$ are in
contrast with the observed suppression, and different $\mgg$
dependence, of $\rocroc$ production \cite{ARGUS} with respect to
$\roro$ \cite{TASSO,PLUTO} in the data for $\q \approx 0$ and $\mgg <
2 \GeV$.

The observed behaviour of $\rho$ pair-production at large momentum
transfer is well described by the QCD-based model developed in
Reference \citen{QCD}, as shown by the analysis of the L3 data
presented in Reference~\citen{ANIKIN}.  On the other hand, $\rho$
pair-production by quasi-real photons is still not well understood,
despite a wide range of theoretical
models~\cite{ATTEMPTS,rosner}. Thus, the study of the $\q$ evolution
of $\rho$ pair-production between these two $\q$ regimes is an
important task in the experimental investigation of vector meson
pair-production in two-photon interactions. This Letter presents
results on the measurement of the process:
\begin{equation}
\label{eq:eqn01}
\EE \to \EE \gamgam \to  \EE \roro
\end{equation}
in a kinematic region of intermediate values of the squared momentum transfer:
\begin{equation}
\label{eq:rangeq2}
0.2 \GeV^2  <  \q  <0.85\GeV^2  
\end{equation}
and for an invariant mass of the hadronic system, $\mgg$, in the interval:
\begin{equation}
\label{eq:rangewgg}
1.1 \GeV  <   \mgg   <3 \GeV .
\end{equation}

The data sample used was collected by the L3 detector\cite{L3} at LEP
at centre-of-mass energies $183 \GeV < \sqrt{s} < 209 \GeV$ and
corresponds to an integrated luminosity of $684.8~\pb$.  Scattered
beam electrons\footnote{Throughout this Letter, the term ``electron''
denotes both electrons and positrons.}  which have radiated photons
with virtualities in the range (\ref{eq:rangeq2}) can be detected
(``tagged'') by the Very Small Angle Tagger (VSAT) \cite{thesistasja}.
The VSAT is an electromagnetic calorimeter made of BGO crystals
installed around the beam line on opposite sides of the L3 detector,
at 8.05 m from the interaction point.  Its geometrical acceptance
covers the polar angle range 5 ${\rm mrad}<\theta<10$ mrad, for
azimuthal angles in the ranges~ $-$1.25 rad$\,<\!  \phi\! <\,$1.25 rad
and \mbox{$\pi-$1.25 rad$\,<\! \phi\!  <\! \pi$+1.25 rad}.
When the electron with the largest scattering angle is detected by the
VSAT, the maximum virtuality of the two photons, $\q$, is, to good
approximation, equal to the transverse momentum squared, $\ptt $, of
the final state hadron system:
\begin{equation}
\label{eq:eqn02}
\q = 2 E_b E_s (1 - \cos \theta_s) \approx E_b E_s \theta_s^2 \approx  \ptt  \,,
\end{equation}
where $ E_b$ is the beam energy, and $ E_s $ and $\theta_s $ are the
energy and the scattering angle of the tagged electron,
respectively. The VSAT provides a means to ensure selection of
exclusive final states by correlating the scattered electron and the detected
hadron system.

The $\roro$ production cross section is determined as a function of
$\mgg$ and $\q$.  The results are compared to the generalised vector
dominance model (GVDM)~\cite{GVDM}.  A measurement of  process
(\ref{eq:eqn01}) in a similar kinematic region was performed at lower
centre-of-mass energy by the PLUTO Collaboration\cite{PLUTO}. The
present measurement represents a tenfold increase of the statistics
compared to that measurement.

\section {Event Selection}

The reaction (\ref{eq:eqn01}), contributing to the process
\begin{equation}
\label{eq:eqn06}
\EE \to \mathrm{e^+ e^-_{tag}}\pi^+\pi^-\pi^+\pi^- \,,
\end{equation}
is identified by a scattered electron, $ \mathrm{e_{tag}}$,
detected in the VSAT and four charged pions measured in the tracking
chamber.  These events are collected by two independent track-triggers
\cite{L3T}.  The trigger efficiency is determined from the data
itself, making use of the redundancy of the triggers, and is around
94\%.

Single-tagged events are selected by requiring one electromagnetic
cluster in the VSAT. This cluster must have energy greater than $50\%$
of the beam energy in order to reduce the background and to ensure a
sufficient containment of the electromagnetic shower.

The event candidates must have exactly four tracks with zero total
charge.  All tracks must come from the interaction vertex, have
transverse momentum greater than $100 \MeV{}$ and an energy loss in
the tracking chamber compatible with the pion hypothesis.
 
Events containing muons are removed from the selected data sample.  A
search for secondary vertices is performed and events with
reconstructed neutral kaons are also rejected.  Energy depositions
above 60 \MeV{} in the electromagnetic calorimeter, not associated
with a charged track, are selected as photons.  An event is allowed to
contain no more than one such photon with energy below 300 \MeV{}
which should not exceed 10\% of the total energy of the four-pion
system. Events containing higher-energy photons are discarded.
 
According to equation (\ref{eq:eqn02}), the transverse momentum
squared $\ptt$ of the four-pion system is used to measure the $\q$ of
the event. It is required to be in the range (\ref{eq:rangeq2}).  For
an exclusive final state, the projections of the momentum vectors of
the electron tag and the four-pion system on to the plane
perpendicular to the beam direction must be back-to-back.  Therefore,
the acoplanarity angle, $\phi_{aco}$, calculated from the difference
between the azimuthal angles of the tagged electron and the four-pion
system, is required to be less than 0.4 rad, as shown in
Figure~\ref{fig:fig1}a.

After all cuts, 1958 events are observed. Their four-pion mass
spectrum is shown in Figure~\ref{fig:fig2}a. The region
(\ref{eq:rangewgg}) is populated by 1836 events, which are used for
the cross section determination.  The mass distribution of the $\pipi$
combinations of the selected events, displayed in
Figure~\ref{fig:fig2}b, shows a strong $\ro$ signal.  A prominent
clustering of entries is observed at the crossing of the $\ro$ mass
bands in the correlation plot of the masses of the neutral $\pipi$
combinations, shown in Figure~\ref{fig:fig2}c. No such resonance
structure is observed in the correlation plot of the masses of the
$\pi^+\pi^+$ and $\pi^-\pi^-$ combinations, presented in
Figure~\ref{fig:fig2}d.  These features of the two-particle mass
correlations give evidence for a signal from $\roro$ intermediate
states.

We also inspect the two- and three-pion mass distributions in the data
for production of higher-mass resonances. The only statistically
significant evidence is for production of the $f_2(1270)$ resonance,
which appears in the two-pion mass spectra in the intervals $2.1 \GeV
< \mgg < 2.5 \GeV$ and $2.5 \GeV < \mgg < 3 \GeV$ as illustrated in
Figure~\ref{fig:figf2}.  Measurement of $f_2(1270)$ production is
beyond the scope of the present study, which is concentrated on $\ro$
pair-production.

\section {Monte Carlo  Modelling and Studies}

To estimate the number of $\roro$ events in the selected four-pion data sample,
we consider non-interfering contributions from the processes:
\begin{eqnarray}
\label{eq:eqn04}
&&  \gamgam \to \roro \,;    \nonumber \\
&&  \gamgam \to \ro \pipi \,;  \\
&&  \gamgam \to \pipi \pipi \,,\, \mathrm{non \! - \! resonant.} \nonumber
\end{eqnarray}
To take into account $f_2(1270)$ production in the region $\mgg > 2.1
\GeV$, we also consider contributions from the processes:
\begin{eqnarray}
\label{eq:eqn03}
&&  \gamgam \to f_2 f_2\,;    \nonumber \\
&&  \gamgam \to f_2 \ro \,;  \\
&&  \gamgam \to f_2 \pipi  . \nonumber 
\end{eqnarray}
Monte Carlo samples of processes (\ref{eq:eqn04}) and (\ref{eq:eqn03})
are generated with the EGPC~\cite{LINDE} program. About 4 million
events are produced for each of the processes (\ref{eq:eqn04}), about
3 million events for the first of the processes (\ref{eq:eqn03}) and
1.6 million events for the two remaining processes.  The $\mgg$ and
$\q$ dependence are those of the $\gamma\gamma$ luminosity
function~\cite{BUDNEV} and only isotropic production and phase-space
decays are included.  The generated events are passed through the full
L3 detector simulation using the GEANT \cite{geant} and GEISHA
\cite{geisha} programs and then processed in the same way as the data,
reproducing the detector behaviour as monitored in the different
data-taking periods.  The scattered electrons are propagated
from the interaction point to the VSAT taking into account
the influence of the magnetic field of the L3 solenoid and the LEP
quadrupole magnets installed between the L3 detector and the
VSAT~\cite{thesistasja}. 

For acceptance calculations,  Monte Carlo events are assigned a
$\q$-dependent weight, evaluated using the GVDM
form-factor~\cite{GVDM} for both photons.  The detection efficiencies
of process (\ref{eq:eqn01}), calculated taking into account the
detector acceptance and the efficiency of the selection procedure, are
in the range of $2\% - 4\%$ and are listed in Tables~\ref{tbl:xsectq2}
and~\ref{tbl:xsectwgg} in different $\q$ and $\mgg$ intervals. 
The efficiency is mostly limited by the kinematics of the two-photon
reaction which boosts the hadronic system along the beam direction,
often resulting in low-angle tracks outside the fiducial tracking
volume. This geometric acceptance is then further reduced by the
limited angular coverage of the VSAT.
 The
detection efficiencies for the other subprocesses from
(\ref{eq:eqn04}) are of the same magnitude as the $\roro$ one and
follow a similar evolution with $\q$ and $\mgg$.  Including the
$f_2(1270)$ branching fraction into two charged pions, the detection
efficiencies for the $\gamgam \to f_2 \ro $ and $ \gamgam \to f_2
\pipi $ processes are of the order of 2\% and the detection efficiency
of the $\gamgam \to f_2 f_2 $ process is about 1.2\%.

For Monte Carlo events passing the selection, the generated energy of
the tagged electron always exceeds $90\% $ of the beam energy, with an
average $<E_{s} / E_{b} >$ = 0.987. This ensures that the
approximation of $\q$ by $\ptt $, given by relation
(\ref{eq:eqn02}), is valid within 1\% in the region
(\ref{eq:rangeq2}).  The $\q$ resolution is determined by the
measurement of $\ptt $ and varies between $8\%$ and $10\%$; the
resolution on $\mgg$ is better than $3\%$.

\section {Background Estimation}

The contribution to the selected sample from $\EE$ annihilation is
negligible. Using 2 million Monte Carlo events of the reaction
$\EE\rightarrow\EE\tau^+\tau^-$ generated with the program
LEP4F~\cite{LEP4F}, the background contribution from this process is
estimated to be $0.6\pm 0.3$ events and is neglected. The background is
mainly due to partially reconstructed events from two-photon
interactions with higher particle multiplicities in the final state,
when tracks or photons escape detection.  Another background
contribution arises from ``fake tags'', {\it i.e.} random coincidences
with off-momentum beam electrons, which give a signal in the
VSAT. These signals correspond to energy depositions comparable with
the beam energy, and are thus not removed by the cut on the energy of
the VSAT cluster.

To estimate the background due to feed-down from higher-multiplicity
final states, we select a data sample of doubly-charged four-pion
events, $\pi^+ \pi^+ \pi^+ \pi^-$ and $\pi^+ \pi^- \pi^- \pi^-$, in
which at least two charged particles are undetected.  In addition, we
also select $\pipi\pipi\piz$ events which are used to account for
background events with undetected neutral pions.

\begin{figure} [p]
  \begin{center}
     \mbox{\epsfig{file=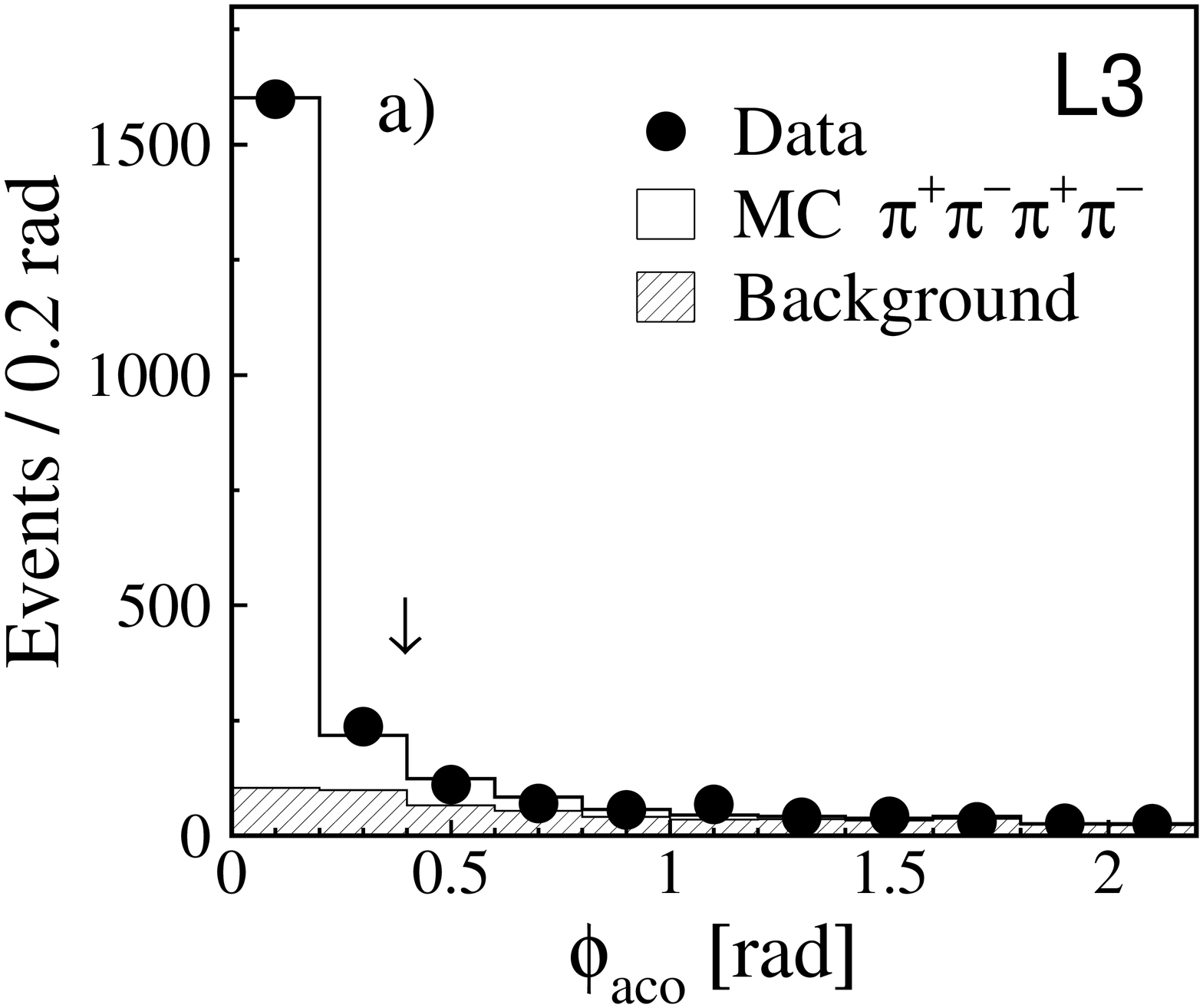,width=0.45\textwidth}}\\
     \mbox{\epsfig{file=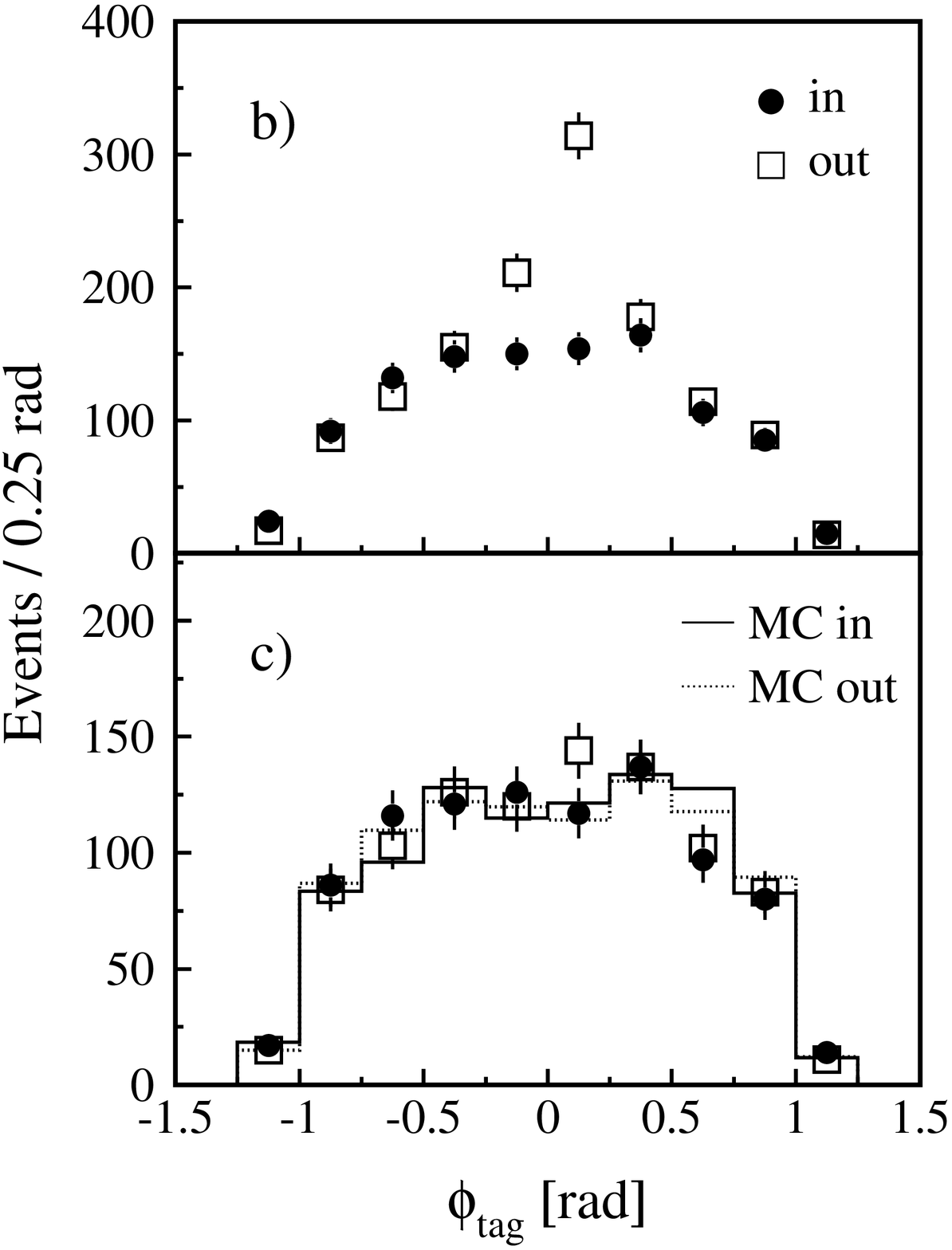,width=0.45\textwidth}}
  \end{center}
  \caption[]{
             (a) Distribution of the acoplanarity angle, $\phi_{aco}$,
             between the electron and the $\pipi\pipi$ system
             for data (points) compared to the four-pion Monte Carlo
             (open histogram) and the background estimated from the data (hatched histogram).
             The arrow indicates the selection cut. The shapes of the
             Monte Carlo and the background are fixed, and their sum
             is normalised to the total number of events.
             (b) and (c) Distributions of the azimuthal angle of the
             tagged electron in the selected events, $\phi_{tag}$, for
             electrons in the inner side of the LEP ring (in) and,
             folded over it, distributions for electrons in the outer side of
             the LEP ring (out).  In (b) all cuts but the
             acoplanarity cut are applied and in (c) all cuts are
             applied and the corresponding four-pion Monte Carlo
             distributions are also shown. 
            }
\label{fig:fig1}
\end{figure}

\begin{figure}[p]
\begin{center}
\vskip -1.5cm
{\epsfig{file=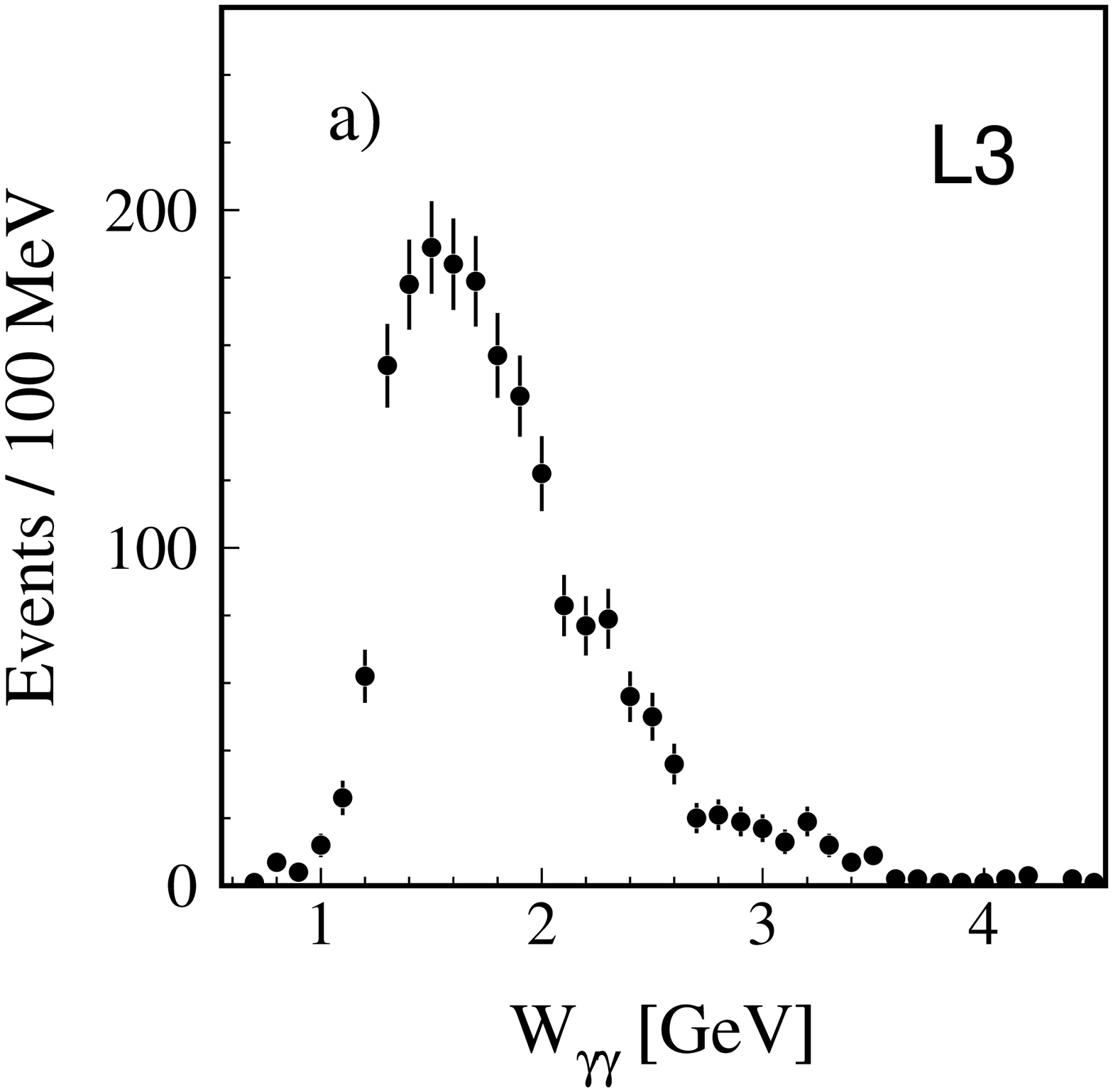,width=0.49\linewidth}}
{\epsfig{file=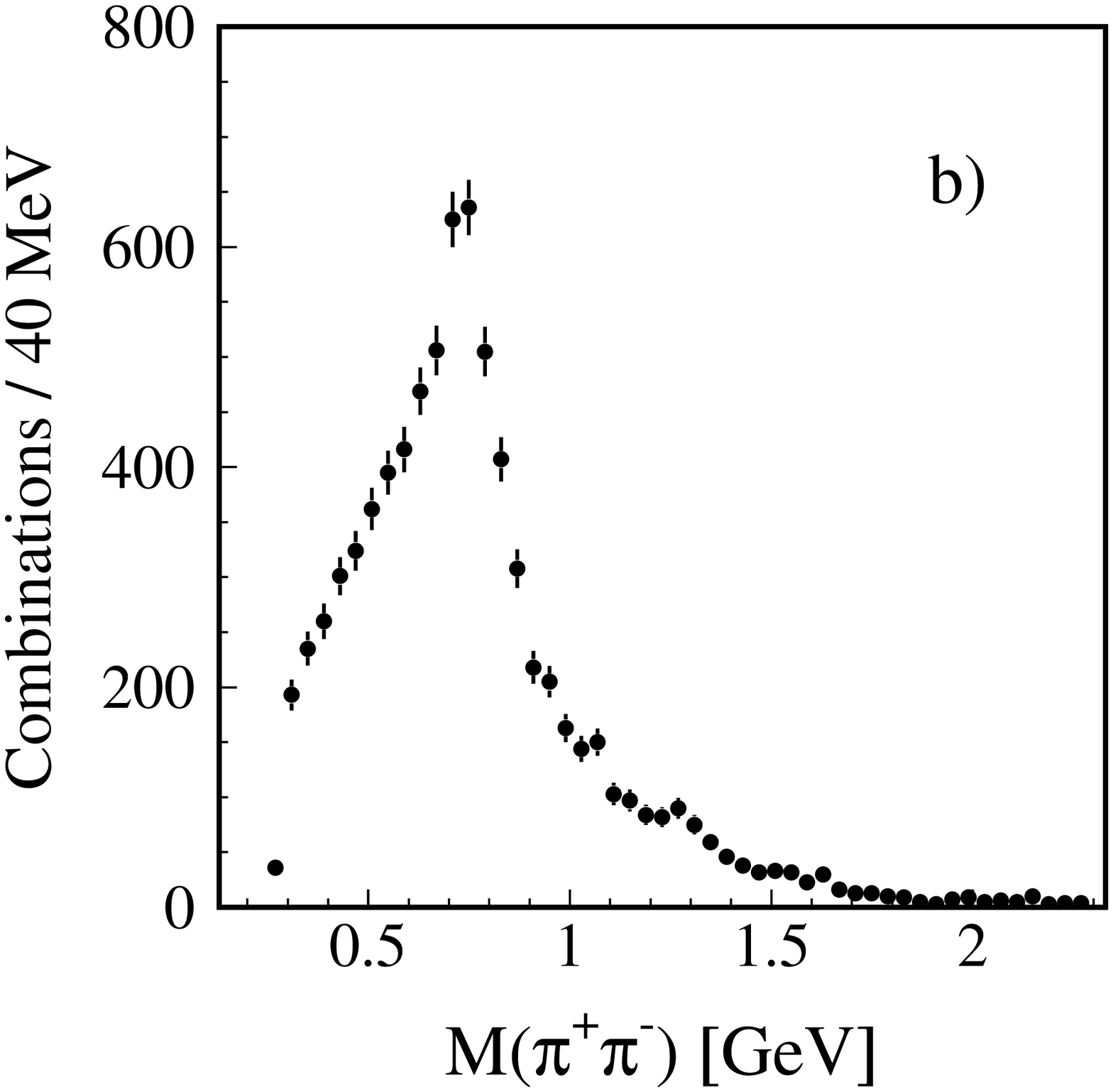,width=0.49\linewidth}}
{\epsfig{file=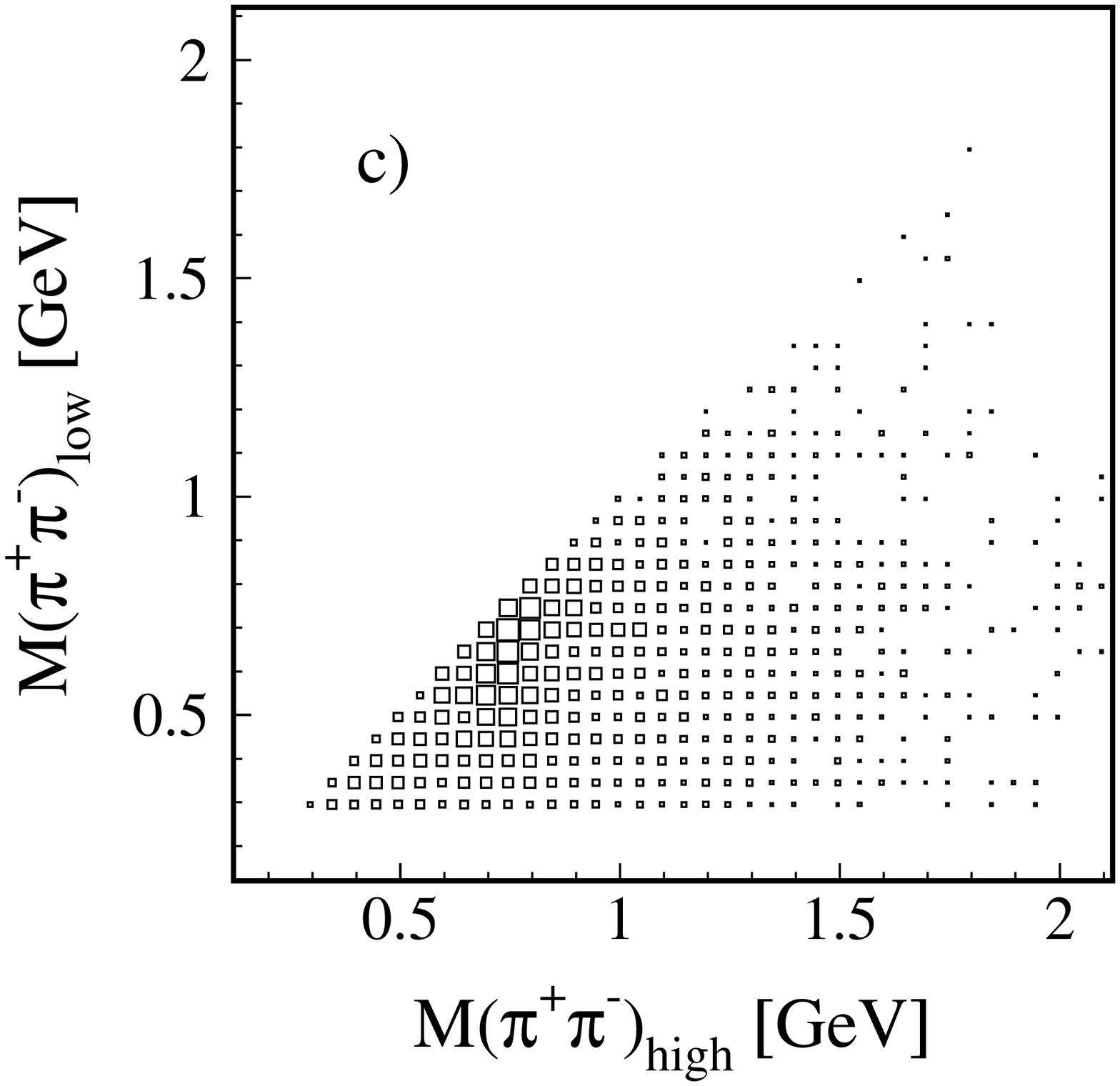,width=0.49\linewidth}}
 {\epsfig{file=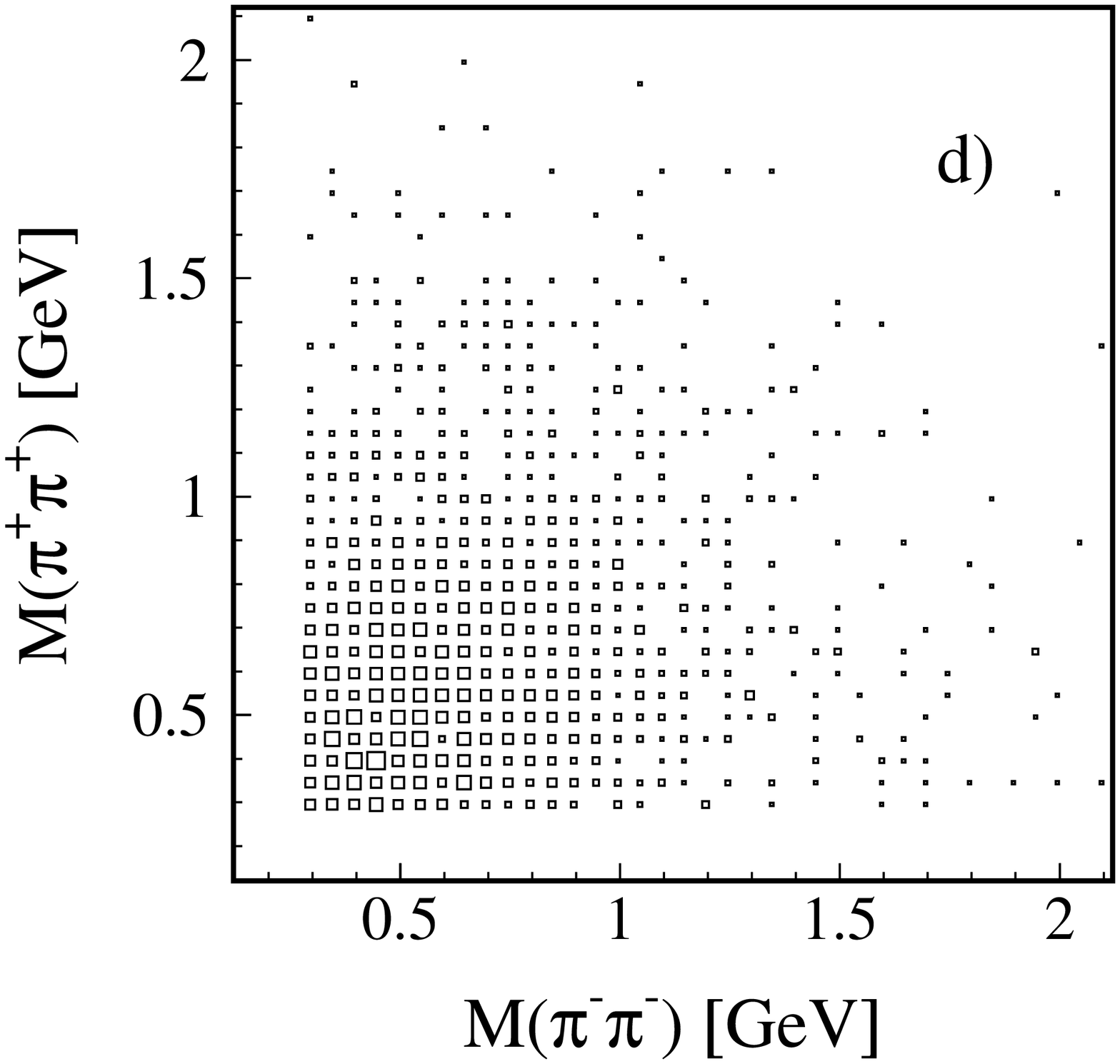,width=0.49\linewidth}}
\end{center}

  \caption[]{Mass distributions for the selected events.
             (a) Mass of the four-pion system, $\mgg$.
             (b) Mass of $\pipi$ combinations (four entries per event).
             (c) Correlation between the  lower versus higher mass
              combinations of the $\pipi$ pairs (two entries per event).
             (d) Correlation between the masses of the $\pi^+\pi^+$ and  $\pi^-\pi^-$ pairs.
             The two-dimensional plots in (c) and (d) have a bin width of
             $50 \times 50 \MeV^2$  and the size of the squares is
             proportional to the number of entries.
           }
\label{fig:fig2}
\end{figure}

All these events are required to pass the event selection procedure,
releasing the charge-conservation requirement for the doubly-charged
events and considering only the $\pipi\pipi$ subsystem of the
$\pipi\pipi\piz$ events.  The  $\phi_{aco}$
distributions of the accepted back\-ground-like data events are
combined with the distribution of selected $\pipi\pipi$ Monte Carlo
events so as to reproduce the $\phi_{aco}$ distribution observed in
data.  
The result of this procedure, applied for
the events in the kinematic region defined by (\ref{eq:rangeq2}) and
(\ref{eq:rangewgg}), is shown in Figure~\ref{fig:fig1}. The resulting
background levels are quoted in Tables~\ref{tbl:xsectq2}
and~\ref{tbl:xsectwgg}.

Dedicated studies show that the off-momentum beam particles at the
VSAT location are dominantly on the outer side of the LEP ring.
Therefore, the related background would appear as an excess in the
number of events having a tag on the outer side of the accelerator
ring, $N_{out}$, with respect to the inner side, $N_{in}$.  This
feature is observed, for instance, in data when the cut on
$\phi_{aco}$ is released, as shown in Figure~\ref{fig:fig1}b. In the
selected data, displayed in Figure~\ref{fig:fig1}c, the ratio
$N_{out}/N_{in} = 1.02 \pm 0.05$ is close to unity, indicating that the
background from fake tags is small.  The ratio of the number of
selected events having tag in the forward versus backward directions
along the beam line, $1.04 \pm 0.05$, is also compatible with unity.  We
note that since the two back\-ground-like data samples used in the
background estimation originate from real physics processes, they
contain a fraction of events with fake tags and take into account the
effect of this background.

\section {Fit Method}

In order to determine the differential $\roro$ production rate, a
maximum likelihood fit of the data to the sum of the processes
(\ref{eq:eqn04}) and (\ref{eq:eqn03}) is performed in intervals of
$\q$ and $\mgg$.

The parameter set, $\Omega$, comprising the six two-pion masses in an event,
namely the four neutral combinations $\pipi$ and the two
doubly-charged combinations $\pi^\pm \pi^\pm$, provides a complete
kinematic description of a four-pion event in our model of isotropic
production and decay.  For each data event, $i$, with measured
variables $\Omega_i$, we calculate the probabilities, $
P_j(\Omega_i)$, that the event resulted from the $j$-th production
mechanisms of the six possible ones as listed in (\ref{eq:eqn04}) and
(\ref{eq:eqn03}).  A likelihood function is defined as:
\begin{equation}
\label{eq:eqn05}
\Lambda = \prod_{i} \sum_{j=1}^{6} \lambda_j P_j (\Omega_i)
 \,,\;\;\;\;\;\; \;\;\sum_{j=1}^{6} \lambda_j = 1,
\end{equation}
where the fit parameter $\lambda_j$ is the fraction of process $j$ in the
$\pipi\pipi$ sample for a given $\q$ or $\mgg$ bin and the product
runs over all data events in that bin.  The probabilities $P_j$ are
determined by the six-fold differential cross sections of the
corresponding process, using Monte Carlo samples and a box
method\cite{BOXMETHOD}.

In the fits we assume that the processes (\ref{eq:eqn03}) involving
$f_2(1270)$ production contribute only for $\mgg > 2.1 \GeV$, as
suggested by the spectra in Figure~\ref{fig:figf2}.  We find the
$f_2$ content in the data to be well described by the $f_2 \pipi$ and
$f_2 \ro$ contributions only.  Therefore, in order to reduce the
correlations between the fitted parameters and their uncertainties, we
exclude the process $ \gamgam \to f_2 f_2$ from further consideration.  Thus,
we perform a five-parameter fit in the $\q$ bins and in the
$\mgg$ bins for the region $\mgg > 2.1 \GeV$, whereas the fits in the
$\mgg$ bins for $\mgg < 2.1 \GeV$ have three parameters and take into
account only contributions from the processes (\ref{eq:eqn04}).

As a check of the fit method, we find that the maximum-likelihood fit
reproduces the $\roro$ content of Monte Carlo test samples within 
statistical uncertainties. Since the analysis procedure is optimised
for deriving the $\roro$ contribution, in the following only the
$\roro$ content and the sum of the rest of the contributing processes,
denoted as ``other $4\pi$'', are considered.

To check the quality of the fit, the $\pipi$ mass distributions of the
data are compared with those of a mixture of Monte Carlo event samples
from the processes (\ref{eq:eqn04}) and (\ref{eq:eqn03}), in
proportions determined by the fit.  The data and Monte Carlo
distributions are in good agreement over the entire $\q$ and $\mgg$
range. As an example $\pi^+\pi^-$ mass distributions are shown in
Figure~\ref{fig:figf2}.  The Monte 
Carlo production model  also provides a good description of the
measured angular distributions, as shown in Figure~\ref{fig:angles}.

\section {Results}

The cross section, $\mathrm{\Delta \sigma_{ee}}$, of the process $\EE
\to\EE\roro$ is measured in bins of $\q$ and $\mgg$. The results are
listed in Tables~\ref{tbl:xsectq2} and~\ref{tbl:xsectwgg}, together
with the efficiencies and the background fractions.  The statistical
uncertainties, listed in the Tables~\ref{tbl:xsectq2}
and~\ref{tbl:xsectwgg}, are those of the fit.  
The differential cross section $d \sigma_{\mathrm{ee}} / d \q$,
derived from $\mathrm{\Delta \sigma_{ee}}$, is listed in
Table~\ref{tbl:xsectq2}.  When evaluating the differential cross
section, a correction based on the $\q$-dependence of the $\roro$
Monte Carlo sample is applied, so as to assign the cross section value
to the centre of the corresponding $\q$ bin~\cite{BIN}.
 
To evaluate the cross section, $\sigma_{\gamma\gamma}$, of the process
$\gamgam \to \roro$, the integral of the transverse photon luminosity
function, $L_{TT}$, is computed for each $\q$ and $\mgg$ bin using the
program GALUGA \cite{GALUGA}, which performs ${\cal O}(\alpha^4)$ QED
calculations. The cross section $\sigma_{\gamma\gamma}$ is derived
from the measured cross section $\mathrm{\Delta \sigma_{ee}}$ using
the relation $\mathrm{\Delta \sigma_{ee}} = L_{TT}
\sigma_{\gamma\gamma}$.  Thus, $\sigma_{\gamma\gamma}$ represents an
effective cross section containing contributions from both transverse
and longitudinal photon polarisations.  The cross section of the
process $ \gamgam \to \roro$ is listed in Table \ref{tbl:xsectq2} as a
function of $\q$ and in Table~\ref{tbl:xsectwgg} as a function of
$\mgg$.  The sum of the cross sections of the other contributing
processes is also given in Tables~\ref{tbl:xsectq2}
and~\ref{tbl:xsectwgg}.

Several sources of systematic uncertainty are considered. The
contribution of the selection procedure, as estimated by varying the
selection criteria, is in the range $4\% - 8\%$.  Monte Carlo
statistics give a contribution in the range \mbox{$1.5\% - 2.3\%$}.
The variations of the acceptance observed when a $\rho$-pole
form-factor is used instead of a GVDM 
form-factor for re-weighting Monte
Carlo events are in the range \mbox{$1\% - 3\%$} for most of the
kinematic region.  The uncertainties of the trigger efficiency, as
determined from the data, are in the range \mbox{$1.9\% - 4\%$}. In
order to estimate the systematic uncertainty of the fit procedure, the
size and the occupancies of the boxes in the box-fit are varied, as
well as the number of bins in which the data is divided for the fits. In
particular, the fits in $\q$ are performed using only three bins,
which results in the same integrated cross section as in the case of
four $\q$ bins. A contribution of $3\% - 7\%$ is derived. Finally, an
uncertainty of \mbox{$2\% - 4\%$} is associated with the background
determination.

All contributions are added in quadrature to obtain the systematic
uncertainties quoted in Tables~\ref{tbl:xsectq2}
and~\ref{tbl:xsectwgg}.

\section {Discussion}

The cross section of the process $ \gamgam \to \roro$ as a function of $\mgg$
is plotted in Figure~\ref{fig:xsectwgg}a, together with the sum of the cross sections 
of the other contributing processes. The  shoulder in the 
latter is due to the contribution of the 
subprocesses involving  $f_2(1270)$ production.
The measured $ \ro \ro $ cross section shows a broad enhancement at
threshold. Figures~\ref{fig:xsectwgg}b and~\ref{fig:xsectwgg}c compare
the measured cross sections with those measured at high $\q$
\cite{L3paper269}. All cross sections decrease with $\q$ and the
variation with  $\q$ of the $ \gamgam \to \roro$ cross section is more
rapid for low values of $\mgg$.

The measured differential cross section $d \sigma_{\mathrm{ee}} / d
\q$ of the reaction $\EE \to\EE\roro$ is shown in
Figure~\ref{fig:xsectq2}a, together with the high-$\q$ data from
Reference~\citen{L3paper269}.  It is fitted to
a form~\cite{DIEHL} expected from QCD-based
calculations\cite{DIEHLPAP}:
\begin{equation}
\label{eq:eqn11}
{ \mathrm{d} \sigma_{\mathrm{ee}} \over \mathrm{d} \q }\sim \frac{1} { Q^n
(\q + < \mgg >^2)^2} \, ,
\end{equation}
where $n$ is a constant and $< \mgg >$ is the average $\mgg$ value of
1.8 \GeV\ for this measurement. Although this formula is expected to be valid
only for $\q \gg \mgg$, we find it provides a good parametrisation of
the $\q$ evolution of all data in the interval $0.2 \GeV^2 < \q < 30
\GeV^2$, with an exponent $ n = 2.9 \pm 0.1$.  In the fit, which
results in $\chi^2 /d.o.f. = 6.9/10$ and is shown by the line in
Figure~\ref{fig:xsectq2}a, only the statistical uncertainties are
considered.

The measured cross section of the process $ \gamgam \to \roro$ as a
function of $\q$ is shown in Figure~\ref{fig:xsectq2}b, together with
the L3 data for $\roro$ production at high $\q$ \cite{L3paper269} and
the PLUTO measurement for $1 \GeV < \mgg < 3.2 \GeV$ \cite{PLUTO}. The
two data sets agree for $\q > 0.3 \GeV^2$ while for low $\q$ values
the L3 data lie below the PLUTO measurement.  The L3 data is fitted
with a form-factor parametrisation based on the GVDM
model~\cite{GVDM}, which is found to reproduce well the $\q$
dependence of our measurements.  Only the statistical uncertainties
are considered in the fit, which results in $\chi^2 /d.o.f. = 7.5/11$.
Figure~\ref{fig:xsectq2}b also shows the result of a $\rho$-pole
form-factor fit to the PLUTO data, as in reference \citen{PLUTO}.  The
L3 data cannot be described by the steeper fall of the $\rho$-pole
parametrisation.


%
%

\newpage
\typeout{   }     
\typeout{Using author list for paper 287 -  }
\typeout{$Modified: Jul 15 2001 by smele $}
\typeout{!!!!  This should only be used with document option a4p!!!!}
\typeout{   }
%
%
%
%
%
%

\newcount\tutecount  \tutecount=0
\def\tutenum#1{\global\advance\tutecount by 1 \xdef#1{\the\tutecount}}
\def\tute#1{$^{#1}$}
\tutenum\aachen            
\tutenum\nikhef            
\tutenum\mich              
\tutenum\lapp              
\tutenum\basel             
\tutenum\lsu               
\tutenum\beijing           
\tutenum\bologna           
\tutenum\tata              
\tutenum\ne                
\tutenum\bucharest         
\tutenum\budapest          
\tutenum\mit               
\tutenum\panjab            
\tutenum\debrecen          
\tutenum\dublin            
\tutenum\florence          
\tutenum\cern              
\tutenum\wl                
\tutenum\geneva            
\tutenum\hamburg           
\tutenum\hefei             
\tutenum\lausanne          
\tutenum\lyon              
\tutenum\madrid            
\tutenum\florida           
\tutenum\milan             
\tutenum\moscow            
\tutenum\naples            
\tutenum\cyprus            
\tutenum\nymegen           
\tutenum\caltech           
\tutenum\perugia           
\tutenum\peters            
\tutenum\cmu               
\tutenum\potenza           
\tutenum\prince            
\tutenum\riverside         
\tutenum\rome              
\tutenum\salerno           
\tutenum\ucsd              
\tutenum\sofia             
\tutenum\korea             
\tutenum\taiwan            
\tutenum\tsinghua          
\tutenum\purdue            
\tutenum\psinst            
\tutenum\zeuthen           
\tutenum\eth               

{
\parskip=0pt
\noindent
{\bf The L3 Collaboration:}
\ifx\selectfont\undefined
 \baselineskip=10.8pt
 \baselineskip\baselinestretch\baselineskip
 \normalbaselineskip\baselineskip
 \ixpt
\else
 \fontsize{9}{10.8pt}\selectfont
\fi
\medskip
\tolerance=10000
\hbadness=5000
\raggedright
\hsize=162truemm\hoffset=0mm
\def\r{\rlap,}
\noindent

P.Achard\r\tute\geneva\ 
O.Adriani\r\tute{\florence}\ 
M.Aguilar-Benitez\r\tute\madrid\ 
J.Alcaraz\r\tute{\madrid}\ 
G.Alemanni\r\tute\lausanne\
J.Allaby\r\tute\cern\
A.Aloisio\r\tute\naples\ 
M.G.Alviggi\r\tute\naples\
H.Anderhub\r\tute\eth\ 
V.P.Andreev\r\tute{\lsu,\peters}\
% ...
\newpage
\begin{list}{A}{\itemsep=0pt plus 0pt minus 0pt\parsep=0pt plus 0pt minus 0pt
                \topsep=0pt plus 0pt minus 0pt}
\item[\aachen]
 III. Physikalisches Institut, RWTH, D-52056 Aachen, Germany$^{\S}$
\item[\nikhef] National Institute for High Energy Physics, NIKHEF, 
     and University of Amsterdam, NL-1009 DB Amsterdam, The Netherlands
\item[\mich] University of Michigan, Ann Arbor, MI 48109, USA
\item[\lapp] Laboratoire d'Annecy-le-Vieux de Physique des Particules, 
     LAPP,IN2P3-CNRS, BP 110, F-74941 Annecy-le-Vieux CEDEX, France
\item[\basel] Institute of Physics, University of Basel, CH-4056 Basel,
     Switzerland
\item[\lsu] Louisiana State University, Baton Rouge, LA 70803, USA
\item[\beijing] Institute of High Energy Physics, IHEP, 
  100039 Beijing, China$^{\triangle}$ 
\item[\bologna] University of Bologna and INFN-Sezione di Bologna, 
     I-40126 Bologna, Italy
\item[\tata] Tata Institute of Fundamental Research, Mumbai (Bombay) 400 005, India
\item[\ne] Northeastern University, Boston, MA 02115, USA
\item[\bucharest] Institute of Atomic Physics and University of Bucharest,
     R-76900 Bucharest, Romania
\item[\budapest] Central Research Institute for Physics of the 
     Hungarian Academy of Sciences, H-1525 Budapest 114, Hungary$^{\ddag}$
\item[\mit] Massachusetts Institute of Technology, Cambridge, MA 02139, USA
\item[\panjab] Panjab University, Chandigarh 160 014, India
\item[\debrecen] KLTE-ATOMKI, H-4010 Debrecen, Hungary$^\P$
\item[\dublin] Department of Experimental Physics,
  University College Dublin, Belfield, Dublin 4, Ireland
\item[\florence] INFN Sezione di Firenze and University of Florence, 
     I-50125 Florence, Italy
\item[\cern] European Laboratory for Particle Physics, CERN, 
     CH-1211 Geneva 23, Switzerland
\item[\wl] World Laboratory, FBLJA  Project, CH-1211 Geneva 23, Switzerland
\item[\geneva] University of Geneva, CH-1211 Geneva 4, Switzerland
\item[\hamburg] University of Hamburg, D-22761 Hamburg, Germany
\item[\hefei] Chinese University of Science and Technology, USTC,
      Hefei, Anhui 230 029, China$^{\triangle}$
\item[\lausanne] University of Lausanne, CH-1015 Lausanne, Switzerland
\item[\lyon] Institut de Physique Nucl\'eaire de Lyon, 
     IN2P3-CNRS,Universit\'e Claude Bernard, 
     F-69622 Villeurbanne, France
\item[\madrid] Centro de Investigaciones Energ{\'e}ticas, 
     Medioambientales y Tecnol\'ogicas, CIEMAT, E-28040 Madrid,
     Spain${\flat}$ 
\item[\florida] Florida Institute of Technology, Melbourne, FL 32901, USA
\item[\milan] INFN-Sezione di Milano, I-20133 Milan, Italy
\item[\moscow] Institute of Theoretical and Experimental Physics, ITEP, 
     Moscow, Russia
\item[\naples] INFN-Sezione di Napoli and University of Naples, 
     I-80125 Naples, Italy
\item[\cyprus] Department of Physics, University of Cyprus,
     Nicosia, Cyprus
\item[\nymegen] Radboud University and NIKHEF, 
     NL-6525 ED Nijmegen, The Netherlands
\item[\caltech] California Institute of Technology, Pasadena, CA 91125, USA
\item[\perugia] INFN-Sezione di Perugia and Universit\`a Degli 
     Studi di Perugia, I-06100 Perugia, Italy   
\item[\peters] Nuclear Physics Institute, St. Petersburg, Russia
\item[\cmu] Carnegie Mellon University, Pittsburgh, PA 15213, USA
\item[\potenza] INFN-Sezione di Napoli and University of Potenza, 
     I-85100 Potenza, Italy
\item[\prince] Princeton University, Princeton, NJ 08544, USA
\item[\riverside] University of Californa, Riverside, CA 92521, USA
\item[\rome] INFN-Sezione di Roma and University of Rome, ``La Sapienza",
     I-00185 Rome, Italy
\item[\salerno] University and INFN, Salerno, I-84100 Salerno, Italy
\item[\ucsd] University of California, San Diego, CA 92093, USA
\item[\sofia] Bulgarian Academy of Sciences, Central Lab.~of 
     Mechatronics and Instrumentation, BU-1113 Sofia, Bulgaria
\item[\korea]  The Center for High Energy Physics, 
     Kyungpook National University, 702-701 Taegu, Republic of Korea
\item[\taiwan] National Central University, Chung-Li, Taiwan, China
\item[\tsinghua] Department of Physics, National Tsing Hua University,
      Taiwan, China
\item[\purdue] Purdue University, West Lafayette, IN 47907, USA
\item[\psinst] Paul Scherrer Institut, PSI, CH-5232 Villigen, Switzerland
\item[\zeuthen] DESY, D-15738 Zeuthen, Germany
\item[\eth] Eidgen\"ossische Technische Hochschule, ETH Z\"urich,
     CH-8093 Z\"urich, Switzerland
\item[\S]  Supported by the German Bundesministerium 
        f\"ur Bildung, Wissenschaft, Forschung und Technologie.
\item[\ddag] Supported by the Hungarian OTKA fund under contract
numbers T019181, F023259 and T037350.
\item[\P] Also supported by the Hungarian OTKA fund under contract
  number T026178.
\item[$\flat$] Supported also by the Comisi\'on Interministerial de Ciencia y 
        Tecnolog{\'\i}a.
\item[$\sharp$] Also supported by CONICET and Universidad Nacional de La Plata,
        CC 67, 1900 La Plata, Argentina.
\item[$\triangle$] Supported by the National Natural Science
  Foundation of China.
\end{list}
}
\vfill


\newpage

\begin{table}[htb]
\begin{center}
\begin{sideways}
\begin{minipage}[b]{\textheight}
\begin{center}
\begin{tabular}{|c|c|c|c|c|c|c|}
\hline
$ \q$ range &  $\varepsilon$ & $\mathit{Bg}$ & $\Delta \sigma_{ee}$ [ pb ] & $ d\,\sigma_{ee}/d\,\q$ [ pb\,/$\GeV^2 \,]$ & 
$ \sigma_{\gamma\gamma}$ [ nb ]  & $ \sigma_{\gamma\gamma}$ [ nb ] \\
 $[\;\GeV^2 \;\;]$ & [ \% ] & [ \% ] & $\roro$ & $\roro$ &  $\roro$ & other $4\pi$  \\ \hline

0.20 -- 0.28 & 2.4 & 8 & $   12.5\phantom{0} \pm 1.18  \pm 0.79 $ & $ 155\phantom{.0} \pm 15\phantom{.0} \pm 10\phantom{.0} $  & $ 9.65 \pm 0.92 \pm 0.62 $ & $ 15.6 \pm 1.19 \pm 0.90 $ \\ \hline
0.28 -- 0.40 & 3.7 & 9 & $   10.9\phantom{0} \pm 0.90  \pm 0.72 $ & $ \phantom{0}89.5 \pm \phantom{0}7.4 \pm \phantom{0}5.9 $  & $ 8.18 \pm 0.68 \pm 0.54 $ & $ 13.0 \pm 0.89 \pm 0.86 $ \\ \hline
0.40 -- 0.55 & 3.0 & 12& $   \phantom{0}6.37 \pm 0.78  \pm 0.54 $ & $ \phantom{0}42.1 \pm \phantom{0}5.1 \pm \phantom{0}3.6 $  & $ 5.59 \pm 0.68 \pm 0.47 $ & $ 12.7 \pm 0.94 \pm 0.90 $ \\ \hline
0.55 -- 0.85 & 2.0 & 20& $   \phantom{0}6.80 \pm 0.95  \pm 0.83 $ & $ \phantom{0}22.1 \pm \phantom{0}3.1 \pm \phantom{0}2.7 $  & $ 4.63 \pm 0.65 \pm 0.57 $ & $ 7.86 \pm 0.81 \pm 0.79 $ \\ \hline
\end{tabular}

\caption{Detection efficiencies, $\varepsilon$,  background fractions, $\mathit{Bg}$,
         and measured production cross sections 
         of the reactions $ \EE \to \EE \roro$, $\gamgam \to \roro$ 
	        and  of the sum of the
         rest of the contributing processes, other $4\pi$,
         as a function of  $\q$ for $1.1\GeV < \mgg< 3\GeV$. 
         The values of the differential cross sections are   corrected to the centre of each bin.
         The first uncertainties are statistical, the second systematic.
         }
\label{tbl:xsectq2}
\end{center}
\end{minipage}
\end{sideways}
\end{center}
\end{table}

\begin{table*}[ht]
\begin{center}
\begin{tabular}{|c|c|c|c|c|c|}
\hline
$  \mgg$-range & $\varepsilon$&  $\mathit{Bg}$ &  $ \Delta \sigma_{ee}$ [ pb ] & 
$ \sigma_{\gamma\gamma}$ [ nb ]  & $ \sigma_{\gamma\gamma}$ [ nb ] \\
$[\;\GeV \;\;]$ & [ \% ]  & [ \% ] &  $\roro$ &  $\roro$ &  other $4\pi$  \\ \hline
1.10 -- 1.30  & 1.8 & 15 & $ 6.94 \pm 1.08  \pm 0.77  $  & $ 8.05 \pm 1.25 \pm 0.89 $ & $ 7.94 \pm 1.43 \pm 0.86 $ \\ \hline
1.30 -- 1.45  & 2.6 & 12 & $ 6.81 \pm 0.85  \pm 0.58  $  & $ 11.8 \pm 1.48 \pm 1.01 $ & $ 14.3 \pm 1.83 \pm 1.28 $ \\ \hline
1.45 -- 1.60  & 2.8 & 9  & $ 7.07 \pm 0.81  \pm 0.62  $  & $ 13.5 \pm 1.55 \pm 1.19 $ & $ 15.9 \pm 1.83 \pm 1.30 $ \\ \hline
1.60 -- 1.75  & 3.1 & 10 & $ 5.61 \pm 0.70  \pm 0.47  $  & $ 11.8 \pm 1.46 \pm 0.99 $ & $ 16.4 \pm 1.77 \pm 1.24 $ \\ \hline
1.75 -- 1.90  & 3.1 & 10 & $ 3.56 \pm 0.57  \pm 0.36  $  & $ 8.17 \pm 1.32 \pm 0.83 $ & $ 18.1 \pm 1.86 \pm 1.58 $ \\ \hline
1.90 -- 2.10  & 3.1 & 11 & $ 3.37 \pm 0.56  \pm 0.38  $  & $ 6.38 \pm 1.07 \pm 0.71 $ & $ 14.0 \pm 1.40 \pm 1.13 $ \\ \hline
2.10 -- 2.50  & 3.2 & 11 & $ 2.25 \pm 0.44  \pm 0.27  $  & $ 2.48 \pm 0.49 \pm 0.30 $ & $ 14.1 \pm 1.07 \pm 0.96 $ \\ \hline
2.50 -- 3.00  & 3.1 & 11 & $ 0.93 \pm 0.28  \pm 0.12  $  & $ 1.01 \pm 0.31 \pm 0.13 $ & $ 6.85 \pm 0.70 \pm 0.61 $ \\ \hline
\end{tabular}
\caption{Detection efficiencies, $\varepsilon$,  background fractions, $\mathit{Bg}$,
         and measured production cross sections 
         of the reactions $ \EE \to \EE \roro$, $\gamgam \to \roro$ 
	        and  of the sum of the
         rest of the contributing processes, other $4\pi$,
         as a function of  $\mgg$ for $0.2 \GeV^2< \q < 0.85 \GeV^2$.
         The first uncertainties are statistical, the second systematic.
         }
\label{tbl:xsectwgg}
\end{center}
\end{table*}

%
\clearpage
\vfil


\clearpage
\vfil


\clearpage

\hskip -0.5cm
\begin{minipage}{.5cm}
\begin{center}
\large
\rotatebox{90}{Entries / 60 MeV}\\
\normalsize
\end{center}
\end{minipage}\begin{minipage}{16.5cm}

{\epsfig{file=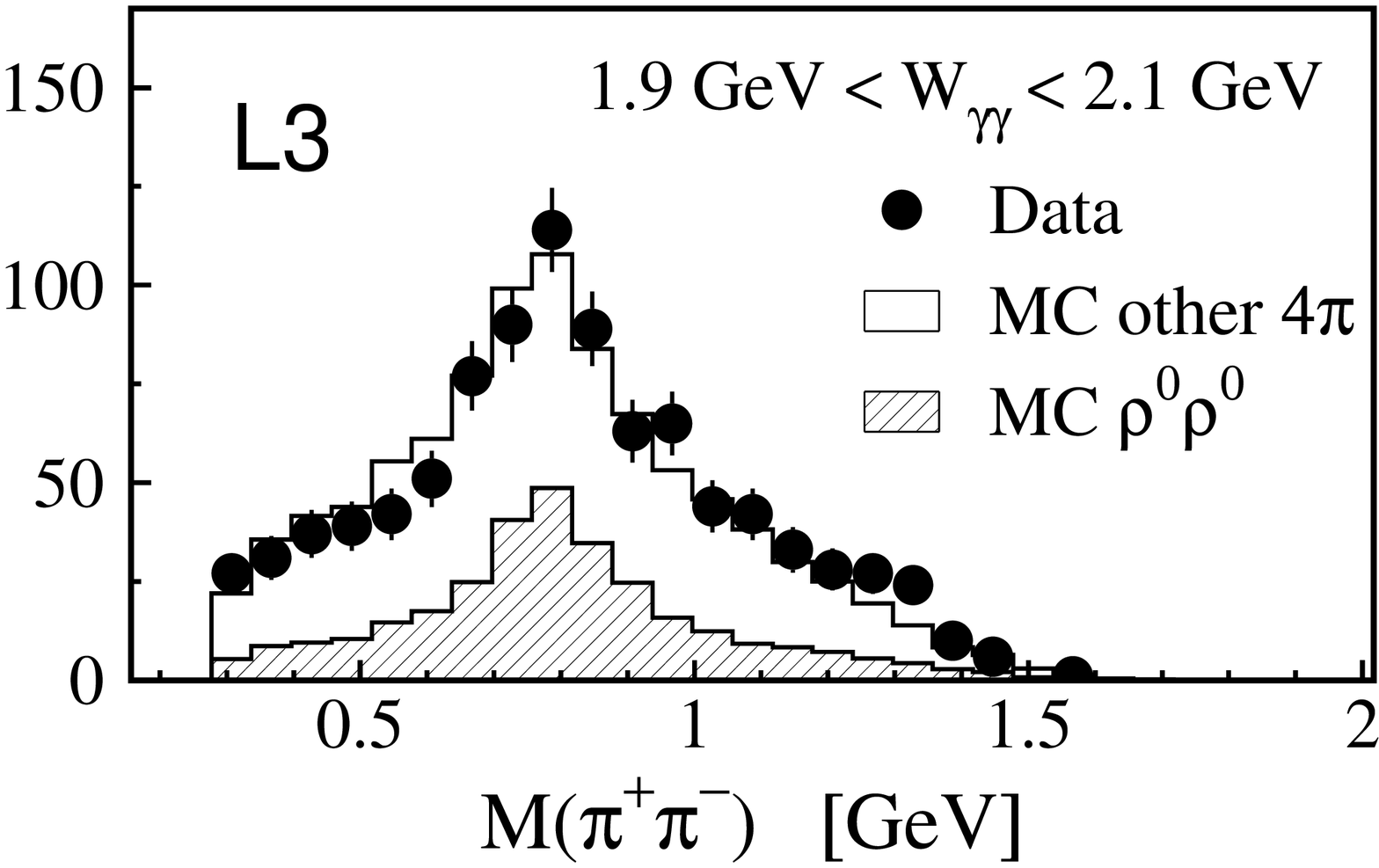,width=0.49\linewidth}}
{\epsfig{file=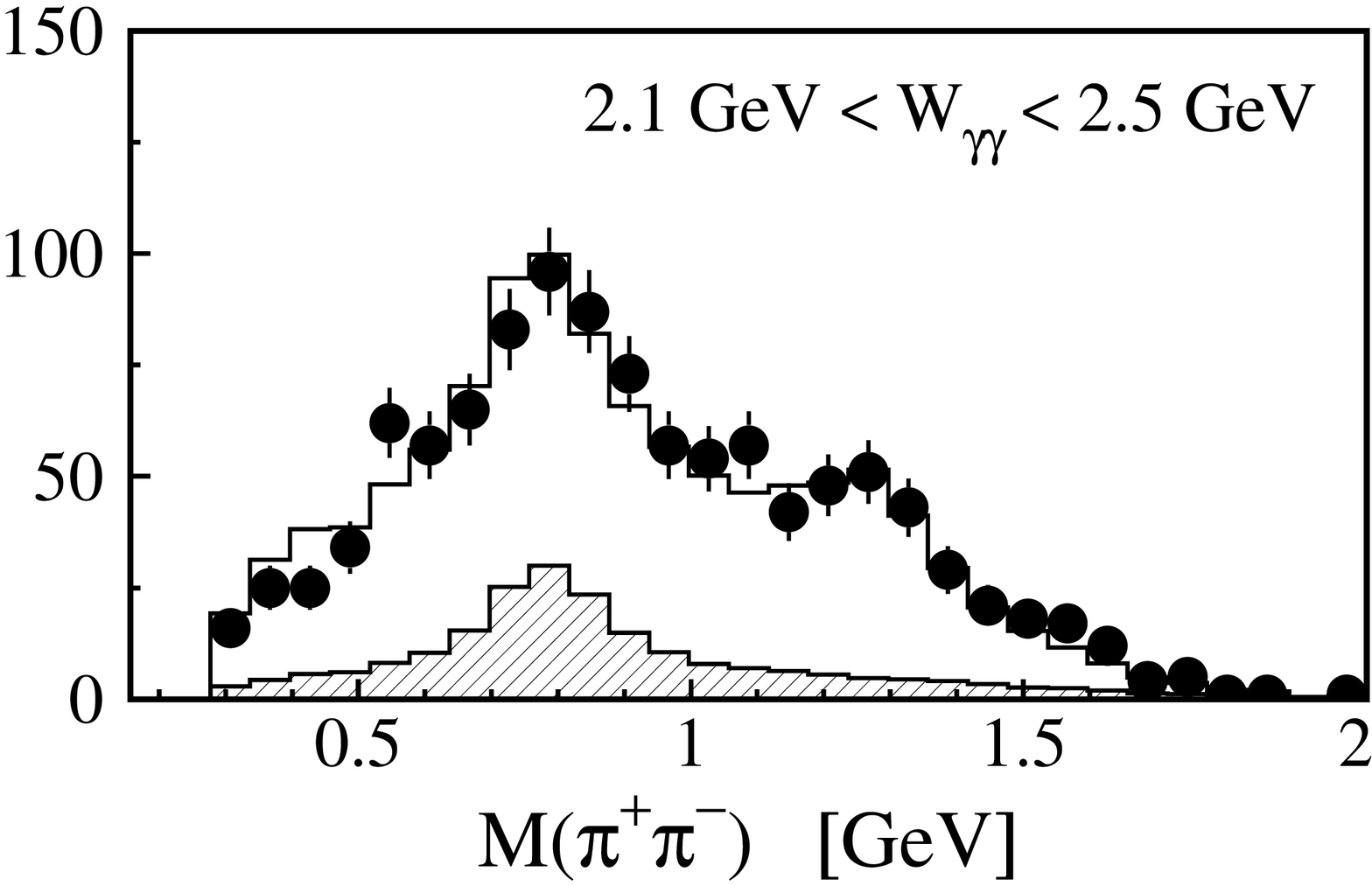,width=0.49\linewidth}}
{\epsfig{file=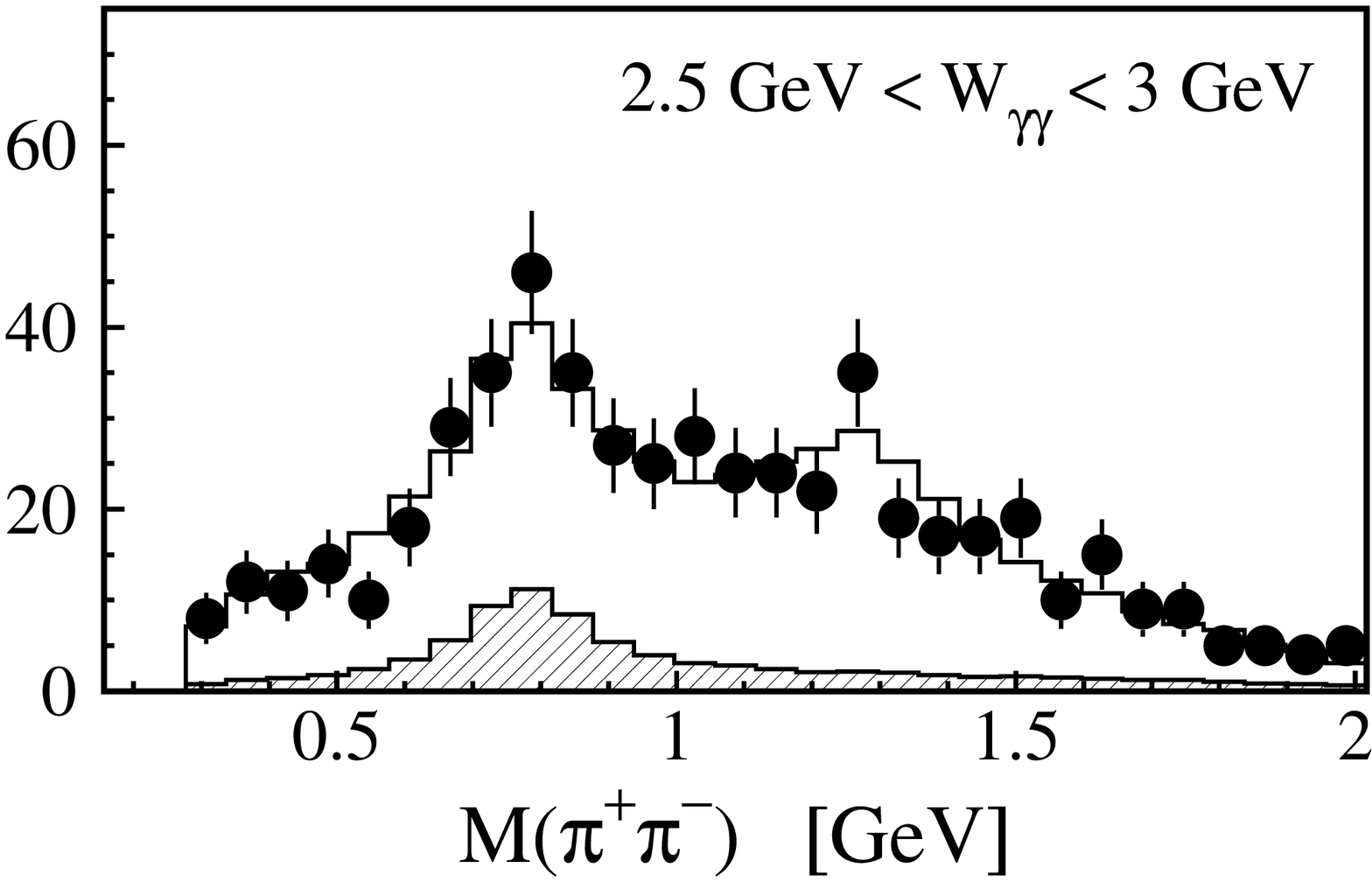,width=0.49\linewidth}}
{\epsfig{file=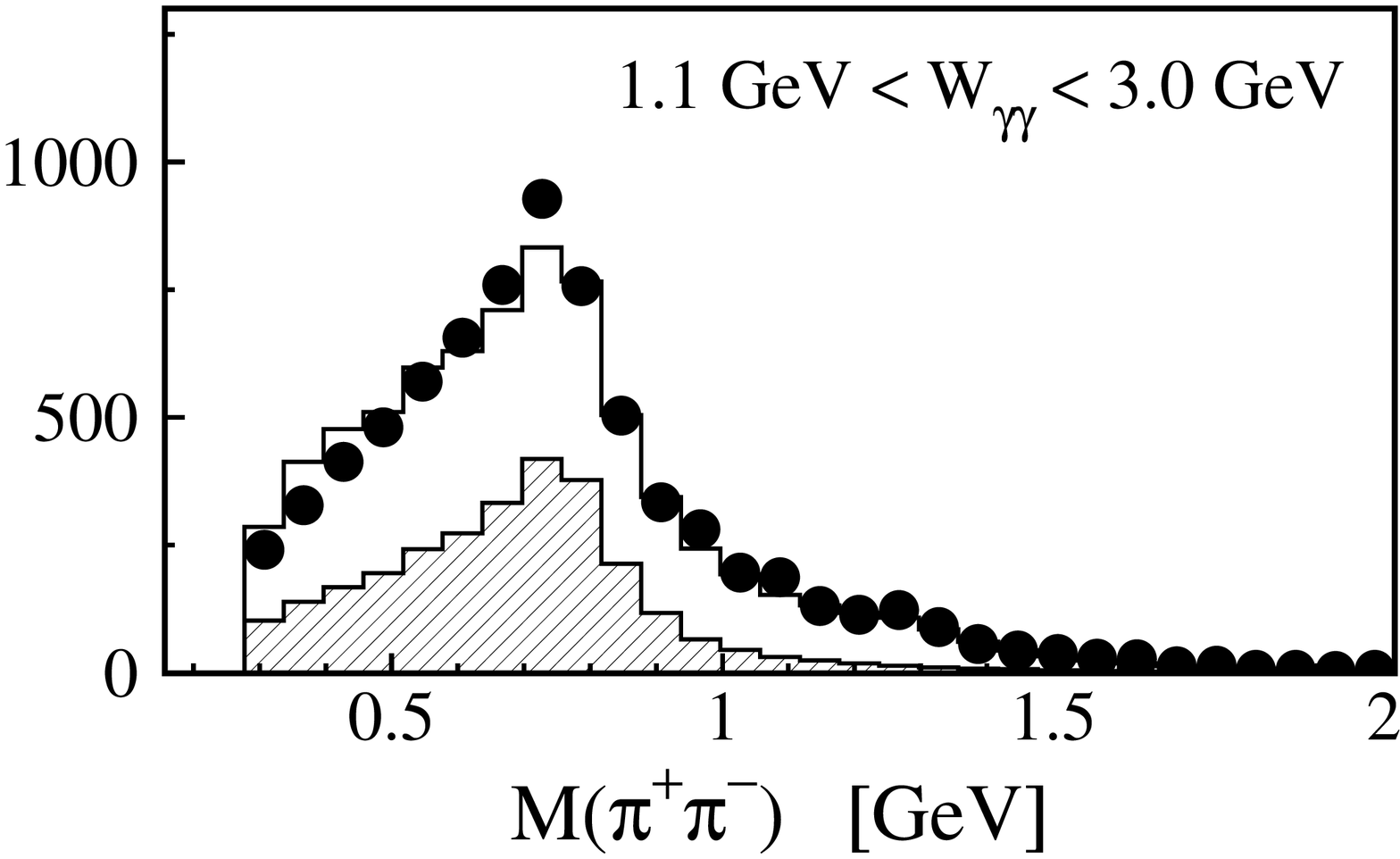,width=0.49\linewidth}}
\end{minipage}

\begin{figure}[ht]
\caption[]{Mass distributions of  $\pipi$ combinations
           (four entries per event) for the three higher $\mgg$
           intervals and for the total sample for  $0.2 \GeV^2< \q <
           0.85 \GeV^2$.  The points represent 
           the data, the hatched area shows the $\roro$ component and
           the open area shows the sum of the rest of the contributing
           processes.  The fraction of the different components are
           determined by the fit and the normalisation is to the total
           number of events. The plot for the entire  $\mgg$ range, $1.1 \GeV < \mgg < 3 \GeV$, is a sum of 
           the distributions of all fitted $\mgg$ intervals. 
           }
\label{fig:figf2}
\end{figure}
\vfil

\clearpage


  \begin{figure} [p]
  \begin{center}
    \vskip -2cm
    \mbox{\epsfig{file=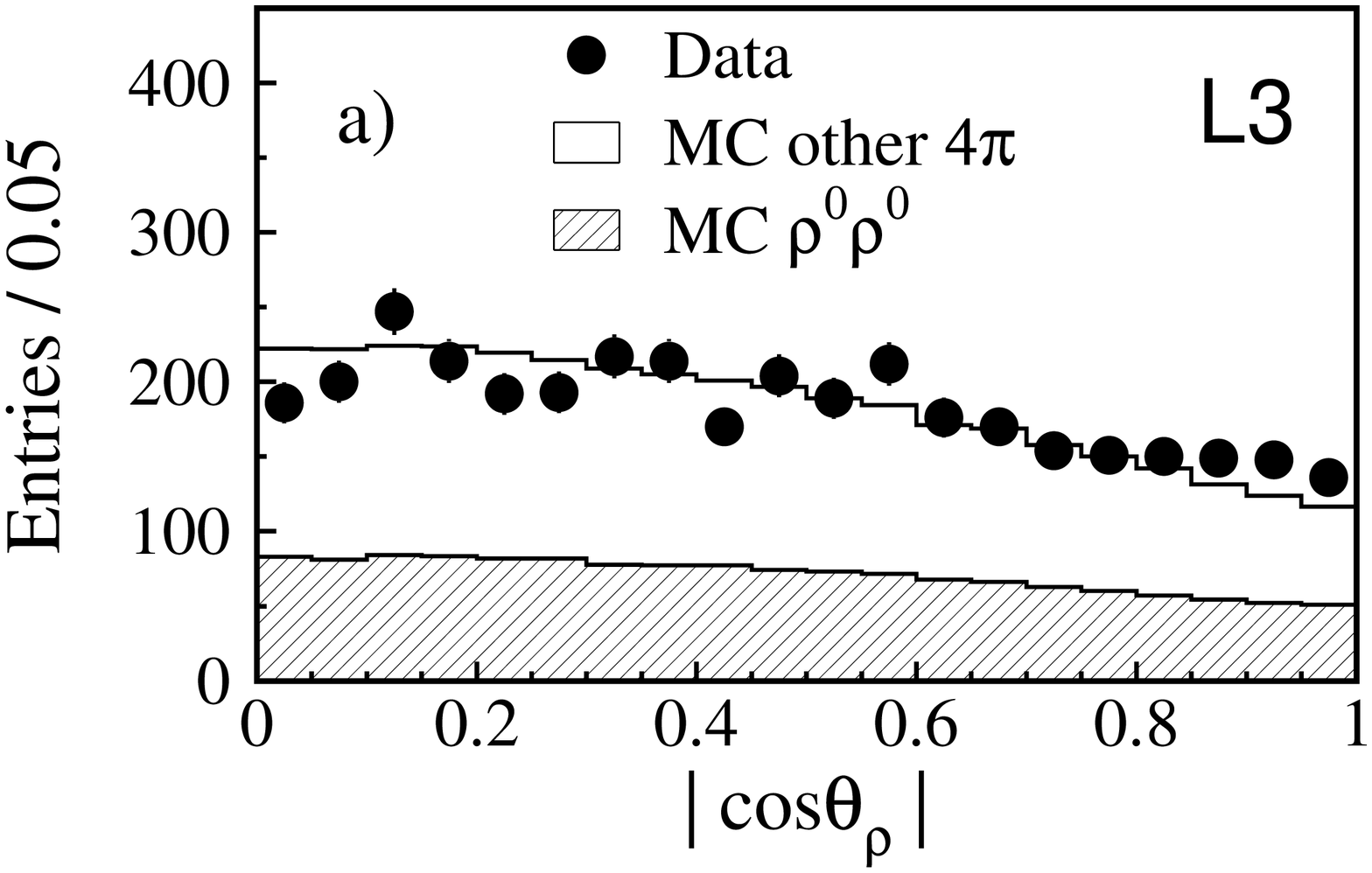,width=0.49\textwidth}}
    \mbox{\epsfig{file=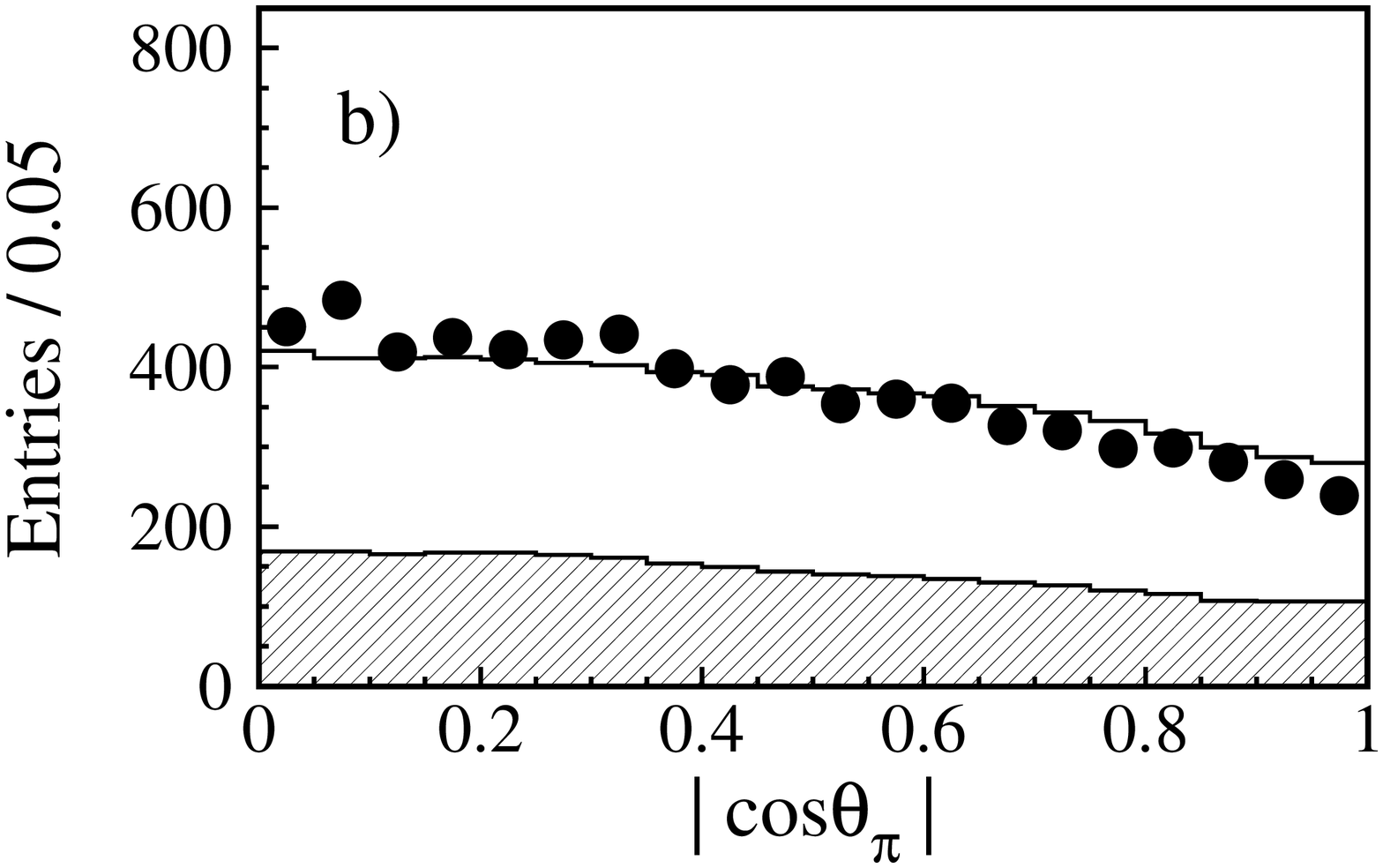,,width=0.49\textwidth}}

    \mbox{\epsfig{file=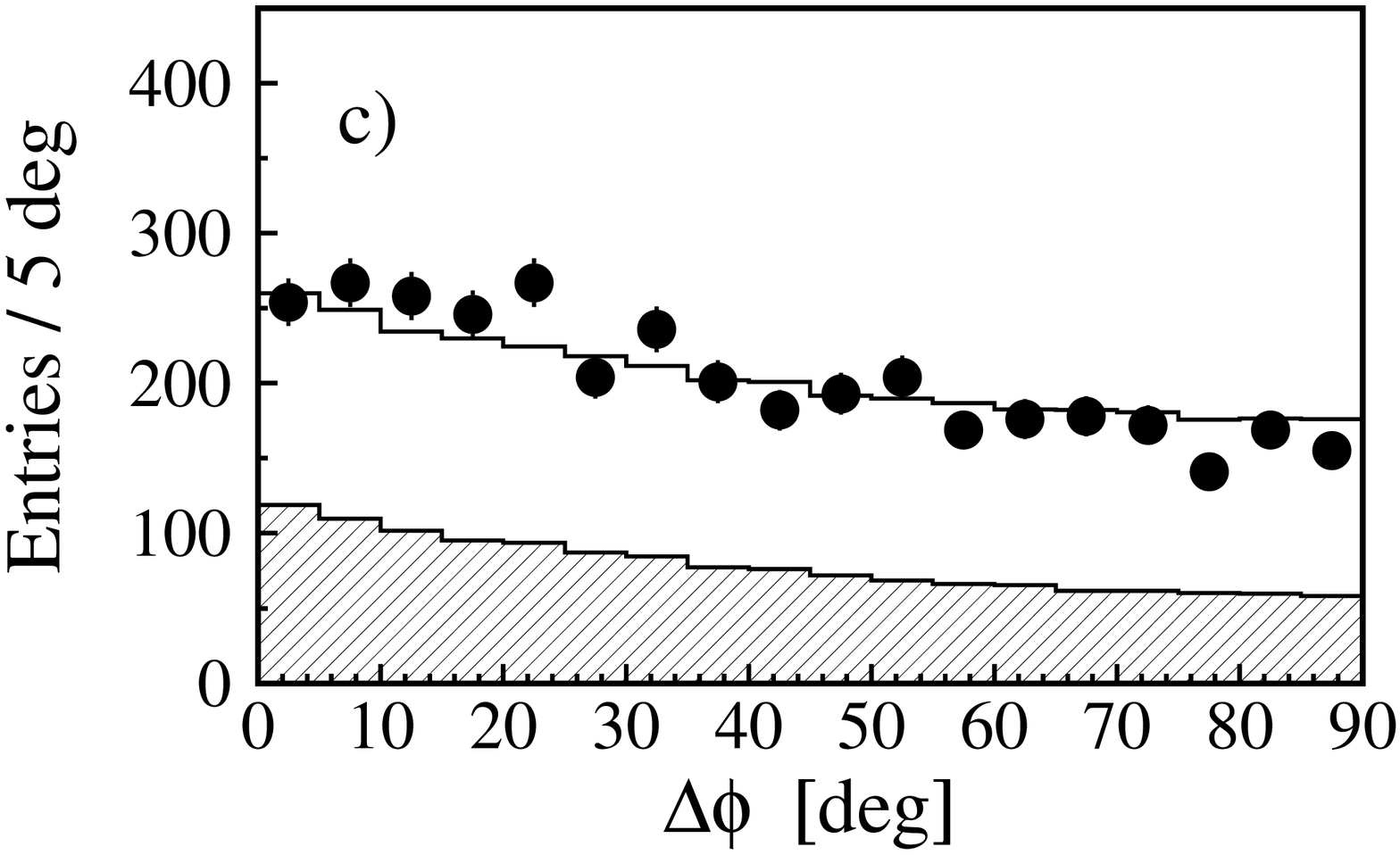,,width=0.49\textwidth}}
    \mbox{\epsfig{file=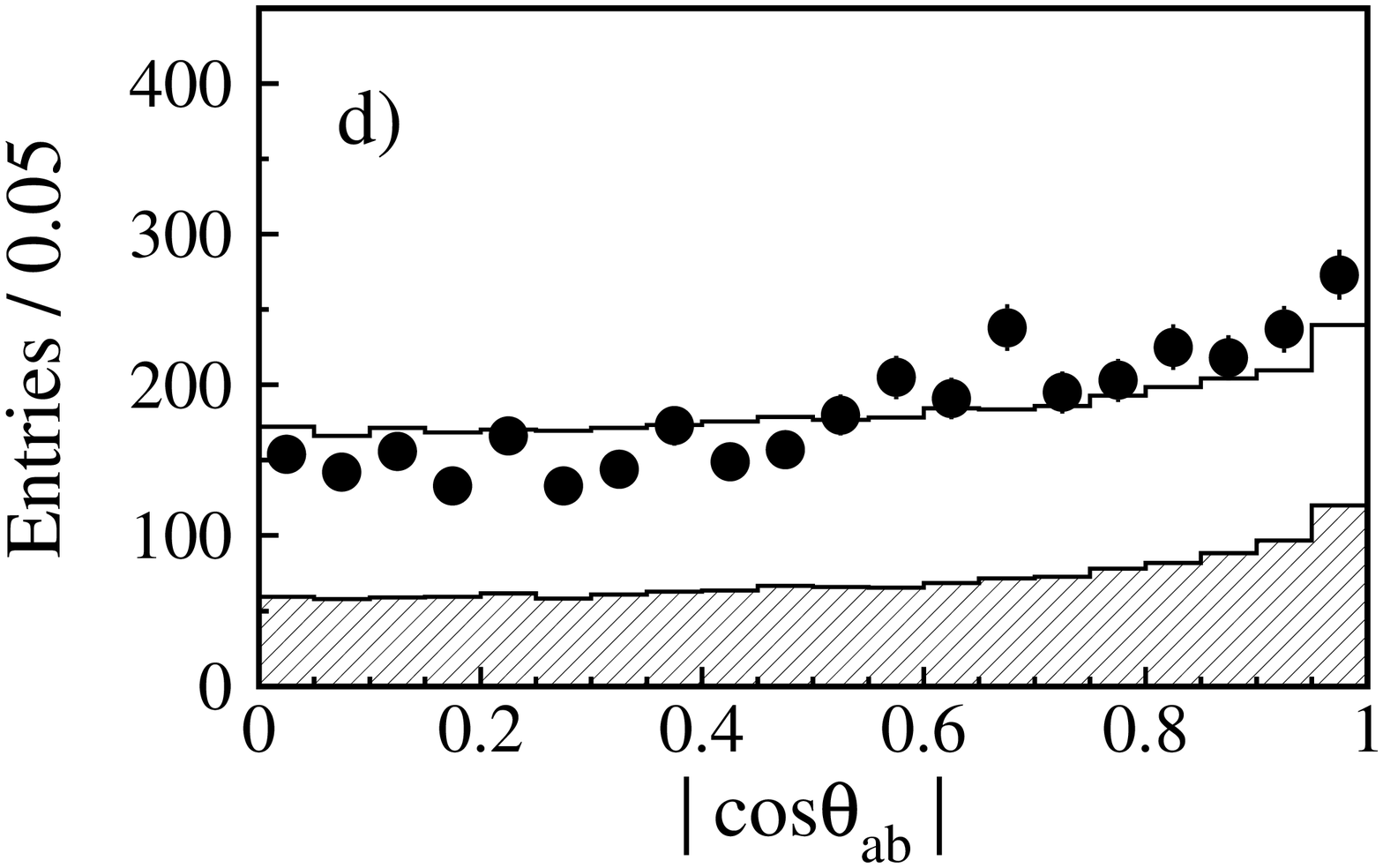,,width=0.49\textwidth}}

  \end{center}
  \caption[]{Comparison of the data   and   Monte Carlo   angular
           distributions for the kinematic regions (1) and (2):
           (a) $\mid \cos \theta_\rho \mid$,  the cosine of the polar angle of
           the $\ro$ with respect to the $\gamgam$ axis in the $\gamgam$
           centre-of-mass system;
           (b) $\mid \cos \theta_\pi \mid$, the cosine of the polar angle of the pion
           in its parent $\ro$ helicity-system;
           (c) $\Delta \phi$, the angle between the decay planes of the two $\ro$ mesons
           in the $\gamgam$ centre-of-mass system;
           (d) $\mid \cos \theta_\mathrm{ab} \mid$, the cosine of the opening angle between
           the two $\pi^+$ directions of flight, each one defined in its parent $\ro$
           helicity-system.
           There are two entries per event in (a), (c) and (d) and four entries per event in (b).
           The points represent data, the hatched area shows  the $\ro\ro$ component
           and the open area shows the sum of 
           the rest of the contributing processes.
           The fraction of the different components are determined by the fit and the
	   normalisation is to the total  number of events.
           }
\label{fig:angles}
\end{figure}
\vfil
\clearpage


 \begin{figure} [p]
  \begin{center}
    \vskip -2cm
    \mbox{\epsfig{file=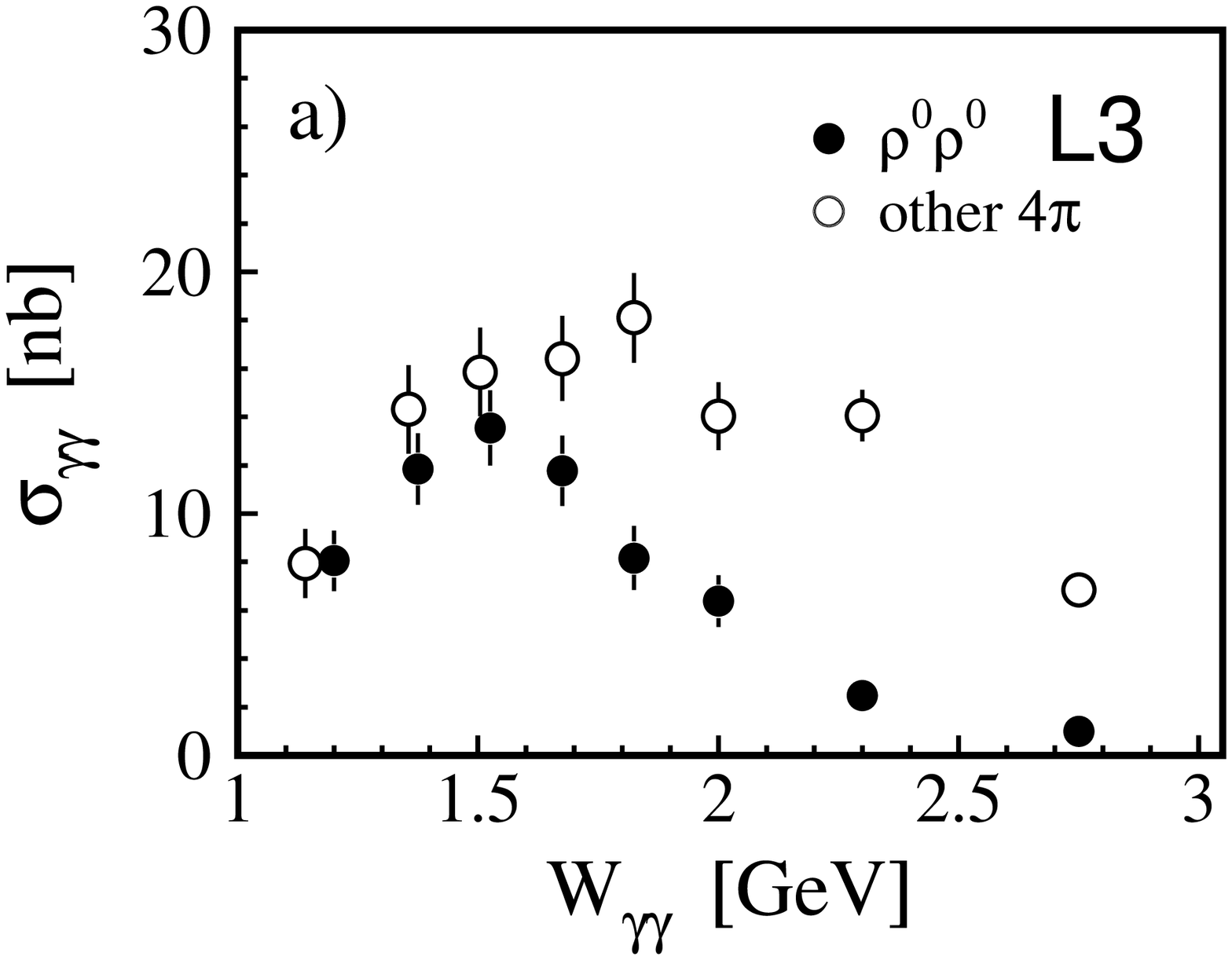,width=0.51\textwidth}}
    \vskip -0.5cm
    \mbox{\epsfig{file=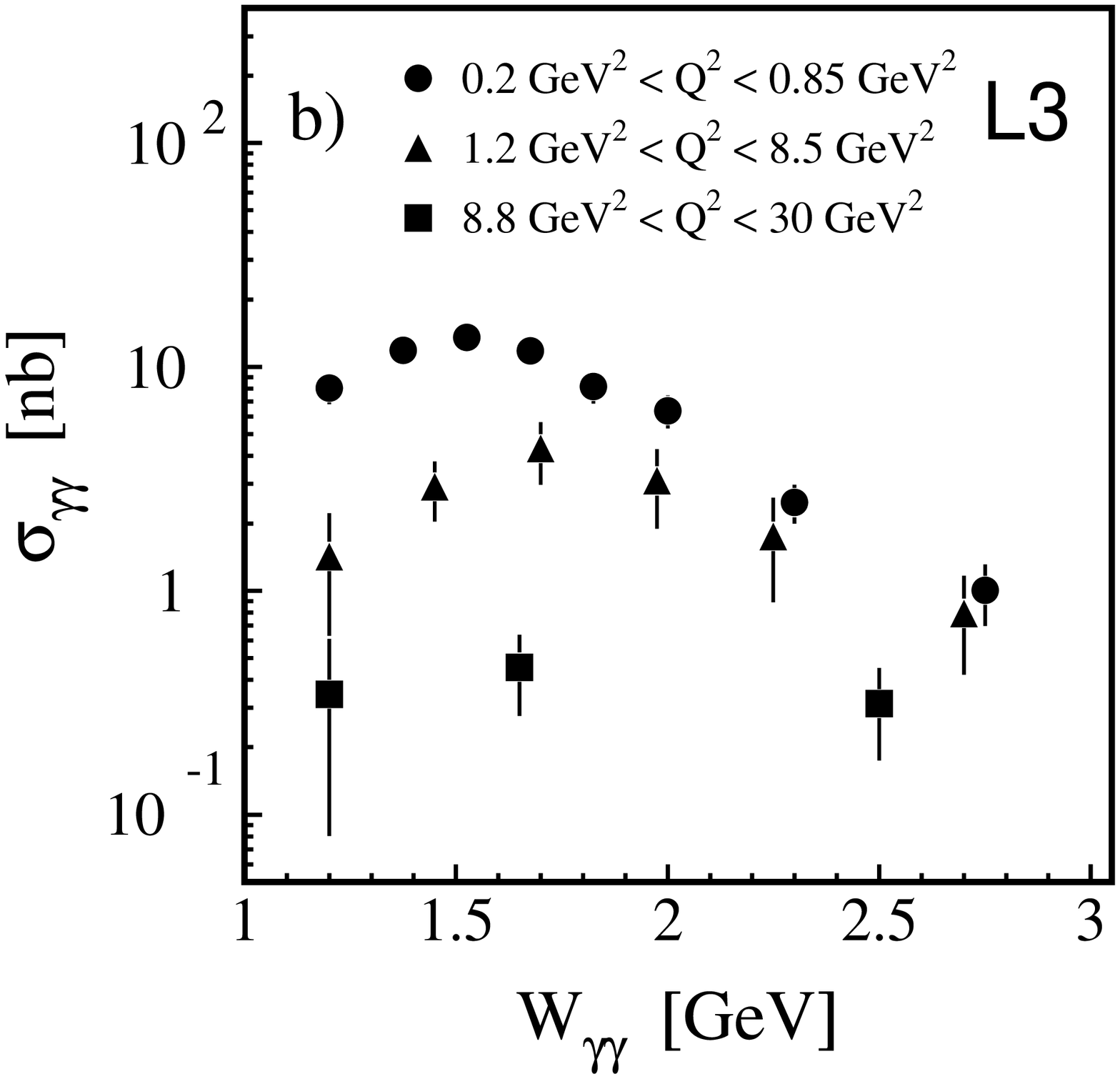,width=0.51\textwidth}}
    \vskip -0.5cm
    \mbox{\epsfig{file=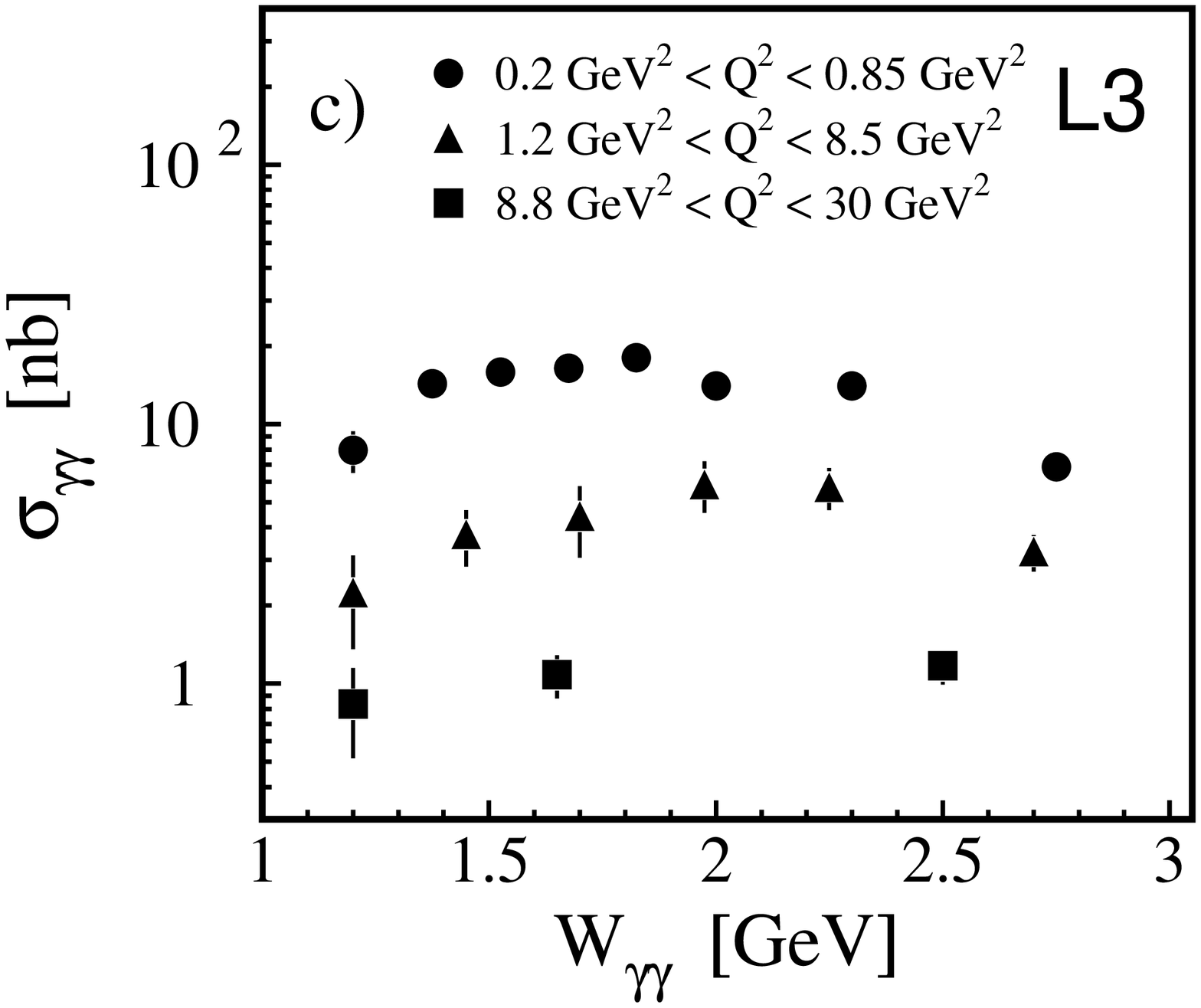,width=0.51\textwidth}}
    \vskip -0.5cm
  \end{center} 
  \caption[]{(a)
            Cross section of the process $\gamgam\to\ro\ro$ (full points)
            and the  sum of the rest of the contributing processes (open points),
            as  a function of  $\mgg$ for  $0.2 \GeV^2 < \q <
           0.85 \GeV^2$.  The  bars show the statistical uncertainties.
            The two sets of points have the same binning and some points are 
            horizontally displaced for better legibility. Comparison
            of the results of this measurement and that at 
	    high $\q$ \protect\cite{L3paper269} for (b) the
            $\gamma\gamma^*\rightarrow\rho^0\rho^0$  
	    process and (c) the sum of the rest of the contributing processes.
           }
\label{fig:xsectwgg}
\end{figure}
%
%

\clearpage
  \begin{figure} [p]
  \begin{center}
    \mbox{\epsfig{file=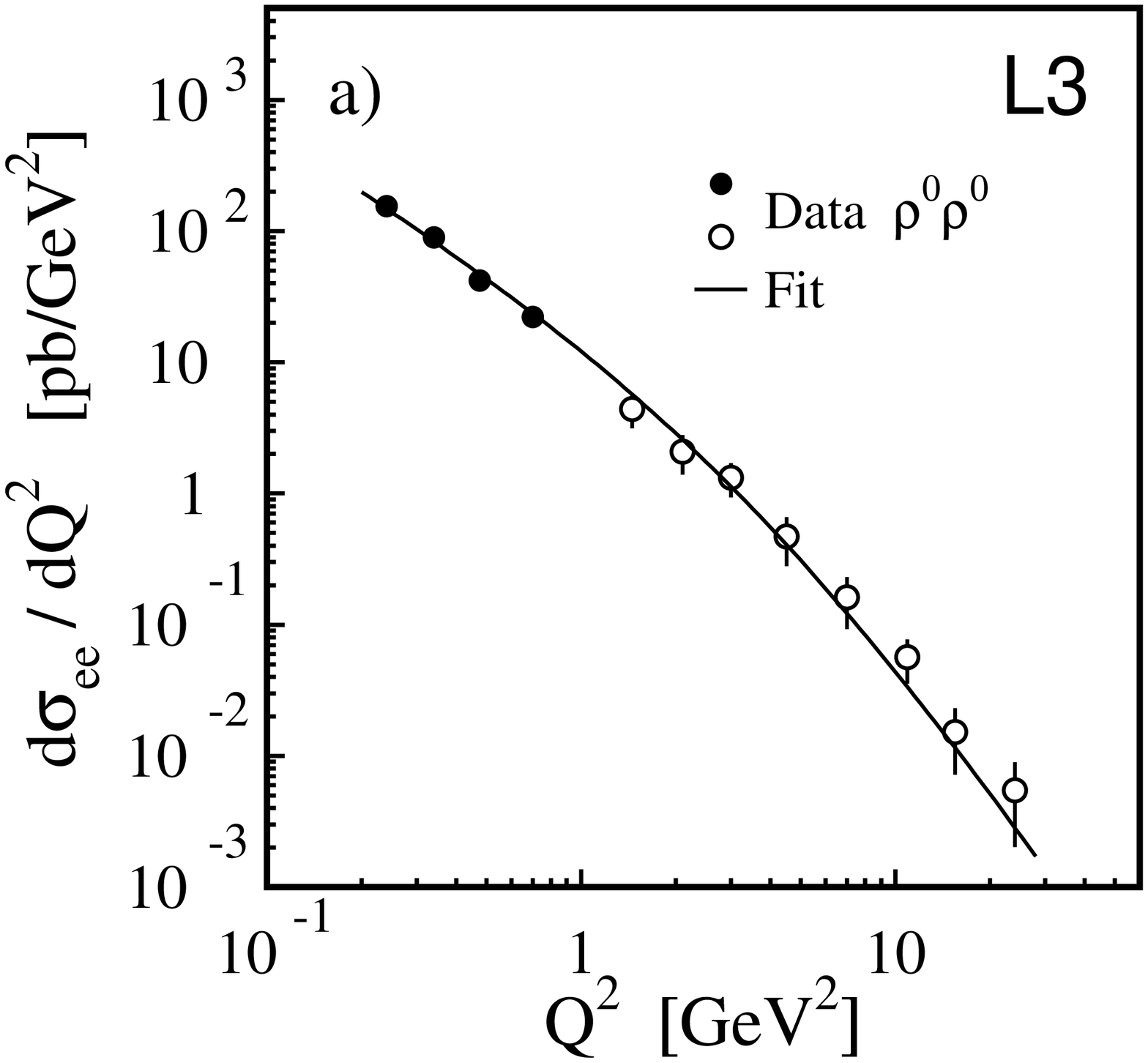,width=0.49\textwidth}}
    \mbox{\epsfig{file=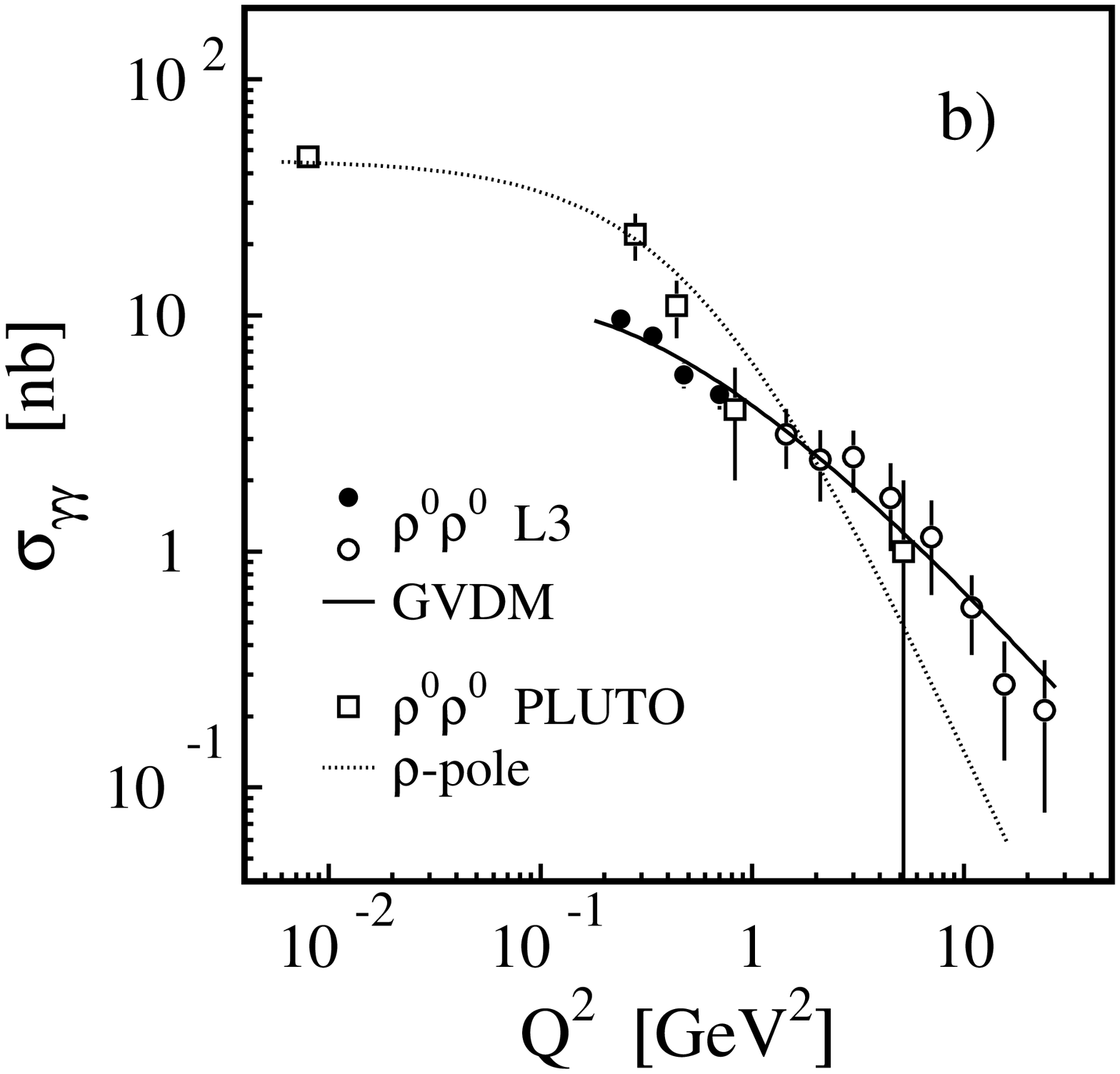,width=0.49\textwidth}}
  \end{center}
 \caption{The $\roro$ production cross section as a function of
         $\q$:
         (a) differential cross section of the process
         $ \EE \to \EE \roro$ and
         (b) cross section of the process $\gamgam \to \roro$.
         The full points show the results from this measurement,
         the open points show the results from the L3 measurement
         of $\ro\ro$ production at high $\q$ \protect\cite{L3paper269} and
         the squares in (b) show the results from the PLUTO
         measurement \protect\cite{PLUTO}.
         The  bars show the statistical uncertainties.
         The L3 measurements are for the interval $1.1 \GeV < \mgg < 3 \GeV$
         and the  PLUTO measurements for $1 \GeV < \mgg < 3.2 \GeV$.
         The line in (a) represents the result of a fit using the
         QCD-inspired form of equation (9).
         The solid line in (b) represents the result of a fit to the L3 data
         based on the GVDM model \protect\cite{GVDM}
         and the dotted line indicates the result of a fit to the PLUTO data
         using a $\rho$-pole form-factor.
           }
\label{fig:xsectq2}
\end{figure}
\vfil

\end{document}